\documentclass[preprintnumbers, prd, onecolumn, floatfix,
preprintnumbers,showpacs, letterpaper, superscriptaddress,nofootinbib]{revtex4}
\usepackage{graphicx,epsfig}
\usepackage{amsmath}
\usepackage{amssymb}
\usepackage{subfigure}
\usepackage{color}
\usepackage{amsfonts}
\begin{document}

\title{Non-singular model universe from a perfect fluid scalar-metric
cosmology}

\author {\textbf{Mohsen Khodadi}}
\email{ m.khodadi@iaufb.ac.ir}
\affiliation{Young Researchers and Elite Club, Firoozkooh Branch,\\ Islamic Azad
University, Firoozkooh, Iran}

\author {\textbf{Kourosh Nozari}}
\email {knozari@umz.ac.ir (Corresponding  Author)}
\affiliation{Department of Physics, Faculty of Basic Sciences,\\
University of Mazandaran, P. O. Box 47416-95447, Babolsar, Iran}
\affiliation{Research Institute for Astronomy and
Astrophysics of Maragha (RIAAM),\\
P. O. Box 55134-441, Maragha, Iran}

\begin{abstract}
To seek for a singularity free model
universe from a perfect fluid scalar-metric
cosmology, we work in the ``\emph{Emergent
Cosmology}" (EC) paradigm which is a non-singular
alternative for cosmological inflation. By using
two methods including Linear Stability Theory and
Effective Potential Formalism, we perform a classical analysis
on the possible static solutions (that are called usually as Einstein
Static Universes (ESU)in literature) in order to study EC paradigm in
a FRW background. Our model contains a
kinetic term of the scalar field minimally coupled to the background geometry without a potential term.
The matter content of the model consists of a perfect fluid plus a cosmological
constant $\Lambda$ as a separate source. In the framework of a local dynamical system analysis,
we show that in the absence or presence of $\Lambda$, depending on some adopted
values for the free parameters of the underlying cosmological
model with flat and non-flat spatial geometries, one gets some
static solutions which are viable under classical linear perturbations. By extending our study
to a global dynamical system analysis, we show that in the presence
of $\Lambda$ with non-flat spatial geometries there is a future global de Sitter attractor in this model.
Following the second method, we derive a new static solution that represents a stable ESU
but this time without dependence on the free parameters of the cosmological
model at hand. As a whole, our analysis suggests the possibility
of graceful realization of a non-singular EC paradigm (i.e. leaving
the initial static phase and entering the inflation period as the
universe is evolving) through either preserving or violation of the
strong energy condition.

\end{abstract}
\pacs {04.20.-q, 04.20.Dw, 98.80.-k}

\maketitle
\section{Introduction}
Cosmic inflation as a paradigm for the early universe
very soon has became a dominant paradigm among cosmologists,
specially by successfully addressing some important unresolved
problems of the standard Big-Bang cosmology (such as the flatness,
horizon and initial seeds of perturbations for large scale structure
formation) \cite{Guth:1981, Starobinsky:1980, Brout:1978, Sato:1981}.
The greatest achievement of this scenario, which so far has been approved
by observational data released with a very remarkable accuracy \cite{Bennet:2003},
is that it could provides a mechanism for producing
a nearly scale invariant spectrum of the primordial density fluctuations.
In fact, through this mechanism it was able for the first time to explain
the origin of inhomogeneities in the universe within the framework of
the causal physics. However, a question then arises is that why despite all
the predictions confirmed by the current observations, still we can not
claim that the cosmological inflation has really happened in
the early stage of the cosmic history. Note that other scenarios
which, as inflation, yield a nearly scale invariant spectrum of the primordial
fluctuations are capable to make predictions consistent with current observational
data of the large scale structure as well as the CMB anisotropies. Therefore,
only matching with observational data cannot be a strict criterion for
the definite acceptance of inflation as a phenomenon occurred in the early
universe. Despite all the amazing phenomenological achievements gained so
far in the context of the inflationary cosmology, it suffers from some non
negligible conceptual issues (for more detailed studies in these issues there
are some seminal review papers such as \cite{Brandenberger:2013, Brandenberger:2014}).
This was a motivation for making a number of alternative (or pre-inflation)
theories such as \emph{``Pre-Big-Bang scenario"} \cite{Gasperini:1993},
\emph{``Matter Bounce cosmology"} \cite{Wand:1999, Finelli:2002},
\emph{``Ekpyrotic Scenario" } \cite{Khoury:2001}, \emph{``Emergent Cosmology"}
\cite{Ellis:2004} and some others\footnote{We emphasize that all
these alternatives to the cosmological inflation must be able to produce a nearly
scale invariant spectrum of the primordial cosmological fluctuations in
one side, and also explain the large scale structure formation of the
current universe from the other side.}. In the present paper, by concerning
on Big-Bang singularity issue, we centralize our attention on the framework of
\emph{``Emergent Cosmology'' (EC)} paradigm. \\

EC is one of the favorite alternatives in recent years in which, by
running the time evolution of the universe from $-\infty$ to $+\infty$
and also by circumventing the Penrose and Hawking singularity theorems
\cite{Hawking:1973}, one manages to provide a non-singular cosmological framework
in contrast to inflationary cosmology models \footnote{Note that the
key assumption of these singularity theorems is the strong energy condition
(ESC) which violates in the inflation scenario.}. According to the
alternative scenarios, by following the evolution of spacetime (assuming
homogeneity and isotropy) towards the past, the growth of the Hubble parameter
$H$ stops in a certain scale and after that $H\rightarrow 0$. In another
words, as the universe moves to the past infinity, the scale factor $a$ also
gets closer to a constant value, $a\rightarrow a_s$.
To be more technical, due to infinity of radius appointed from $H$ (that is,
$R_{H} \sim H^{-1}$) in this quasi-solid initial state (the so called Einstein
static universe (ESU)), the physical wavelength of fluctuations stays almost constant.
So, at the end by falling the $R_{H}$ into a microscopic value, conditions provide the
situation for passing the emergent phase to the standard cosmology framework.
It is importance to note that Ellis et al. suggested this paradigm within the
standard general
relativity (GR) by concentrating on the positive space curvature condition
for the universe. It is noteworthy that this requirement also is supported
observationally since observational data as analyzed by WMAP7 \cite{Komatsu:2011}
address the possibility of a closed universe with 68 percent confidence
level. However, this scenario was defeated at the beginning of its birth.
Since it has been shown that ESU elicited from the basic
gravitational theory (i.e. GR), is not able to survive stable against the prevailing
perturbations in very early moments of the universe and therefore is expected to
decay rapidly \cite{Barrow:2003}. Therefore, this paradigm was developed into
the modified gravity based models (including dynamic equations supported
by some correction terms) with the hope to improve the stability conditions of the ESU
at the high energy regime. For instance in references such as \cite{Canonico:2010}-\cite{Parisi:2012}
that are working within some specific gravity frameworks, the authors have followed this issue.
It is important to note that the EC paradigm (as a non-singular alternative
scenario which is realizable successfully within the framework of any modified gravity),
should satisfy the following two conditions:
\begin{itemize}
  \item \emph{The issue of the existence as well as the stability of ESU against
initial perturbations.}
  \item \emph{Joining to the standard cosmological history by finding a successful
  graceful exit mechanism.}
\end{itemize}
In this regard, some of the above mentioned references just have focused on the first
condition without addressing to the latter one. In this regard, in some of the modified
gravity based cosmological models reported these years there is still no possibility for
deriving a non-singular model universe even from the EC point of view. \\

Due to important role played by scalar fields (from both phenomenological and theoretical
viewpoints) in cosmological models, we study the above mentioned EC paradigm within a FRW
model with a matter part consisting a perfect fluid plus a cosmological constant $\Lambda$
(as a separate source) along with a dimensionless scalar field that is minimally coupled to
the background geometry. To be more specific, in our toy model the dimensionless scalar field
is coupled to the Lorentz invariant measure $\sqrt{-g}d^4x$ constructed from the metric in order
to extract a scalar-metric model universe which results in a non-singular cosmology. We note that
the perfect fluid of our toy model is treated in the Schutz's formalism \cite{Schutz:1970, Shutz:1971}.
In recent years this formalism has been used frequently within different contexts,
see \cite{Alvarenga:2003}-\cite{Khodadi:2016} for instance. The Schutz's formalism provides a
context in which by using some canonical methods \cite{Lapchinskii:1977}, one gets a super
Hamiltonian in order to derive classical
equations of dynamics in the underlying model universe. The rest of the present paper
is devoted to a dynamical analysis of ESU(s) from two viewpoints: the linear stability
theory and effective potential formalism, in the absence or presence of cosmological constant $\Lambda$.
In other words, in the relevant sections we will try to find a well realization of the EC paradigm within a perfect
fluid scalar-metric model universe to get a non-singular cosmological model.
As we will show, both in the presence and absence of the cosmological constant
(and yet with a classical viewpoint of spacetime), there is the chance to get rid
of the initial Big Bang singularity in a fascinating manner.

\section{The equations of motion}

The model universe that we are going to study the issue of EC is a FRW cosmological
model that includes a matter sector with perfect fluid and a cosmological constant
$\Lambda$. In this setup a scalar field minimally interacts with gravity with coupling
to the background metric. To start, by setting the units as $c=\hbar=16\pi G=1$, we
introduce the following total action \footnote{Note that although the first part of this action may looks like as the action for an
inflationary toy model, this is not actually the case since there is no self-interaction potential term, $V(\phi)$, that is necessary in \emph{standard} inflation paradigm.}
\begin{equation}\label{e2-1}
{\cal S}=\int_M d^4x \sqrt{-g}\left[R-g^{\mu \nu}
\phi_{,\mu}\phi_{,\nu}\right]+2\int_{\partial M}d^3x
\sqrt{h}h_{ab}K^{ab}+\int_M d^4 x \sqrt{-g}(p+\Lambda)~,
\end{equation} for the model universe.
In this action $R$, $K^{ab}$ and $h_{ab}$
represent the Ricci scalar, the extrinsic curvature and induced
metric on the three-dimensional spatial hypersurfaces,
respectively. This action corresponds
to a simple form of the scalar-tensor theory introduced by Saez and Ballester (SB)
\cite{SB:1986} in which by minimal coupling of the metric with a dimensionless
scalar field in a simple manner, provides a scalar-metric framework of gravitation
\footnote{We note that there are some different extensions for SB scalar theory. Its simplest extension
includes a non-standard kinetic term via a generic function $F(\phi)\phi_{,\mu}\phi_{,\nu}$
without a scalar potential term \cite{Sabido:2009}. However, in this extension it
can be seen easily that by a re-parametrization of the scalar field as $\phi \rightarrow \tilde{\phi}$,
one can always make $F(\phi)=1$ which means that the generic function is just a gauge quantity.
There are other generalizations of the SB scalar field theory as a toy model of cosmological inflation
in which a potential term is included in the action, see for instance \cite{Nojiri:2006, Capo:2006}.}. Concerning phenomenological
worthy of the SB scalar field theory in classical level, it solves the problem
of missing matter in cosmologies with non-flat spatial geometry and also suggests some quite reliable
descriptions related to the weak fields. We note that all matter content of the theory
is not addressed by the third part in the above action (which contains the pressure of the fluid
$p$ plus a cosmological constant $\Lambda$), but the scalar field component of SB theory also contributes.
Here, unlike the standard inflationary model, the role of the scalar
field in matter budget of the model universe arises just from the standard kinetic term so that
the self-interaction potential term has no contribution.
The second term in the above action may
seems a bit vague. The origin of this boundary term backs
to this point that Ricci scalar in the
gravitational section of the action (\ref{e2-1})
contains second derivatives of the metric tensor,
which is an unconventional trait of field theories.
Of course, with a not so complicated calculation,
one can show that this boundary term will be removed
through variation of $\int_{M} d^{4}x \sqrt{-g}R$.
We are interested in taking the standard equation
of state (EoS) $p=\omega \rho$ as a perfect fluid
with energy density $\rho$. To be more clarified, we give
a little more explanation about the matter term including
the perfect fluid in action (\ref{e2-1}). One, by referring
to \cite{Hawking:1973},  will meet a action
as $S_{P.F}=-\int d^4x\sqrt{-g}\zeta(1+\upsilon)$
(here $\zeta$ and $\upsilon$ are called the density
of fluid's and internal energy, respectively) relevant
to the perfect fluid which is known as Hawking-Ellis
action. There it is shown that by definition of $\rho=\zeta
(1+\upsilon)$ and $p=\zeta^2\frac{d\upsilon}{d\zeta}$, for
the fluid's energy density and pressure, via variation
of the action $S_{P.F}$ in terms of metric, one
gets the standard energy-momentum tensor. Astonishingly,
it has been stated that this result, just with
the same method, can be repeated through taking
the perfect fluid within the Lagrangian, see
\cite{Schutz:1970, Shutz:1971} for more details.
Generally, in the FRW models based Schutz's
formalism, there exists no torsion. Namely,
the fluid's four-velocity can be expressed as
\cite{Schutz:1970, Shutz:1971}
\begin{equation}\label{e2-3}
U_{\nu} = \frac{1}{h} (\epsilon_{,\nu} +
\theta S_{,\nu})~,
\end{equation}in which thermodynamical potentials
$h$ and $S$ denote the specific enthalpy and the specific
entropy, respectively. It is also noteworthy that in Schutz's formalism there
is no explicit physical interpretation for the variables $\epsilon$ and
$\theta$. The above mentioned fluid's four velocity satisfies the
normalization condition $U_{\nu}U^{\nu}=-1$. By taking
the isotropic and homogeneous assumptions in the model
universe at hand, the geometry of spacetime
can be characterized by the FRW metric as
\begin{equation}\label{e2-4}
ds^2 = N^2(t) dt^2 - a^2(t)
\left[ \frac{dr^2}{1-kr^2}+ r^2 d\vartheta^2
+ r^2 \sin^2\vartheta~d\varphi^2 \right]~,
\end{equation}where $N$ and $a(t)$ show
the lapse function and scale factor respectively.
Also, for the curvature constant $k$ there are three
normalized values $+1,0,1$ by concerning a closed,
flat and open universe respectively. The fluid sector
of the action (\ref{e2-1}) can be reexpress via some
thermodynamic equations as follows
\begin{eqnarray}\label{e2-5}
\begin{array}{ll}
\rho = \rho_0(1+ \Pi)~,\\
h = 1 + \Pi + \frac{P}{\rho_0}~,\\
\tau dS = d\Pi + P d \left(\frac{1}{\rho_0}\right)~.
\end{array}
\end{eqnarray}
The quantities $\tau$, $\rho$, $\rho_0$ and $\Pi$ in the
above equations are temperature, total mass
energy density, rest mass density and specific
internal energy respectively. Using these
thermodynamic relations, it is easy to demonstrate
that
\begin{equation}\label{e2-6}
P= \frac{\omega}{(1+\omega)^{\frac{\omega+1}
{\omega}}}h^{\frac{\omega+1}
{\omega}}e^{-\frac{S}{\omega}}~.
\end{equation}
In a comoving system including a perfect fluid
four velocity $U_{\mu} =\left(N,0,0,0\right)$
supported by a cosmological constant $\Lambda$, we
get the following relation
\begin{equation}\label{e2-7}
{\cal L}_{\Lambda} = a^3 N^{-\frac{1}
{\omega}}~\frac{\omega (\dot{\epsilon}+\theta \dot{S})
^{\frac{\omega+1}{\omega}}}{(1+\omega)^{\frac{\omega+1}
{\omega}}}~ e^{-\frac{S}{\omega}} + N a^{3}\Lambda ~,
\end{equation}for the Lagrangian which $h>0$ and
$(\dot{\epsilon}+\theta \dot{S})>0$.
Then, the Hamiltonian of the system is obtained as follows
\footnote{To achieve this form of the Hamiltonian, which
is a function of $P_T$, we have applied the
following canonical transformations \cite{Lapchinskii:1977}
\begin{eqnarray}
\begin{array}{ll}
T=-P_S e^{-S}P_{\epsilon}^{-(\omega+1)} ~~~ \mbox{and}~~~ P_T =
P_{\epsilon}^{\omega+1}
e^S~ \nonumber.
\end{array}
\end{eqnarray} As an advantage, these canonical transformations
let us to follow a dynamical system with
more parameters \cite{Babak:2010, Khodadi:2016}.}
\begin{equation}\label{e2-8}
H_{M-\Lambda} =  \dot{\epsilon}P_{\epsilon}+ \dot{S} P_{S}
- {\cal L}_{M-\Lambda}=
N \frac{P_T}{a^{3\omega}}
- N a^{3}\Lambda~.
\end{equation}
This Hamiltonian is a linear function of $P_T$. Also conjugate momentums,
depending on parameters $\epsilon$ and $S$,
are defined as $P_{\epsilon}=\frac{\partial}{\partial \dot{\epsilon}}{\cal L}_
{\Lambda}$ and $P_{S}=\frac{\partial}{\partial \dot{S}}{\cal L}_{\Lambda}$
respectively. One also can obtain the following point-like expression for the
Lagrangian of the gravitational part of the model at hand
\begin{eqnarray}\label{e2-10}
{\cal L}_{G}=\frac{6a\dot{a}^{2}}{N}+
6kNa-\frac{1}{N}F(\phi)\dot{\phi}^{2}a^{3}~.
\end{eqnarray}
Now, by having the Lagrangian relevant to the matter and gravitational
sectors of the theory, we are able to construct the Hamiltonian for our model
in terms of the conjugate momenta by using the following relation
\begin{equation}\label{e2-11}
H=\dot{a}P_a+\dot{\phi}P_{\phi}+\dot{\epsilon}P_{\epsilon}+\dot{S}P_S-{\cal
L}~,
\end{equation} so that $H= H_{G}+H_{M-\Lambda}$, ${\cal L}={\cal L}_{G}
+{\cal L}_{M-\Lambda}$ and also $P_{q}=\frac{\partial {\cal L}}{\partial
\dot{q}}$ where $q=a,~\phi$. Therefore, the super Hamiltonian relevant to
the minisuperspace of the model universe at hand takes the following form
\begin{equation}\label{e2-12}
H=H_G + H_{M-\Lambda} = N \left(-\frac{1}{24}
\frac{P_a^2}{a} + \frac{1}{4a^{3}}P_{\phi}^{2}-6ka
+\frac{P_T} {a^{3\omega}} - \Lambda a^{3} \right) ~.
\end{equation}
The last two terms including $P_T$ and $\Lambda$ in this
Hamiltonian clearly address the matter part
of the model. In Eq. (\ref{e2-12}), $N$ can be labeled as
the Lagrange multiplier holding the constraint equation
$H=0$. By regarding $\dot{T}=\{T,H\} = \frac{N}{a^{3\omega}}$,
in the case of selecting a gauge as
\begin{equation}\label{e2-13}
N=a^{3\omega}~,
\end{equation} the variable $T$ in Eq. (\ref{e2-12}) takes the role of a
time coordinate (i.e. $T=t$) in the cosmological model at hand. At this point,
for a classical description of the dynamics in this model universe, we present
the following Hamiltonian equations
\begin{eqnarray}\label{e2-14}
\left\{
\begin{array}{ll}
\dot{a}=-\frac{a^{3\omega-1}}{12}P_a~,\\\\
\dot{P_a}=-\frac{a^{3\omega-2}}{24}P_a^2+\frac{3a^{3-3\omega}}
{4F(\phi)a^4}P_{\phi}^2+6ka^{3\omega}+3\omega a^{-1}P_{0}+2\Lambda
a^{3\omega+2}~,\\\\
\dot{\phi}=\frac{a^{3\omega-3}}{2F(\phi)}P_{\phi}~,\\\\
\dot{P_{\phi}}=0~,~~~\Rightarrow P_{\phi}=\mbox{Constant}~,\\\\
\dot{P_{T}}=0~,~~~\Rightarrow P_{T}=\mbox{Constant}~,
\end{array}
\right.
\end{eqnarray} which are actually the same as the classical equations
of motion. By taking time deravitit of the third equation in the above
system, we get
\begin{equation}\label{e2-15}
\frac{\ddot{\phi}}{\dot{\phi}}+3(1-\omega)
\left(\frac{\dot{a}}{a}\right)=0~,
\end{equation} for the field $\phi$. Integrating the second
order equation we arrive at the following expression
\begin{equation}\label{e2-16}
\dot{\phi}^2=C a^{6\omega-6}~,
\end{equation} with a numerical constant as $C>0$.
At the end, in the case of canceling the momenta from the
system (\ref{e2-14}), the differential form of the Hamiltonian
constraint $\cal{H}$=0, becomes
\begin{equation}\label{e2-17}
-6a^{1-3\omega}\dot{a}^2+\dot{\phi}^2a^{3-3\omega}
-6ka^{1+3\omega}+P_{T}-\Lambda^{3\omega+3}=0~,
\end{equation}
where putting (\ref{e2-16}) into this relation the first equation
of motion acquires the following form
\begin{equation}\label{e2-18}
(\frac{\dot{a}}{a})^2= \frac{C}{6}a^{6\omega-6}-ka^{6\omega-2}
+\frac{P_T}{6}a^{6\omega}\rho-\frac{\Lambda}{6} a^{6\omega}~.
\end{equation}
By taking differential of equation (\ref{e2-18}) with respect to time,
the second equation of motion takes the following form
\begin{equation}\label{e2-19}
\frac{\ddot{a}}{a}=\frac{C}{6}a^{6\omega-6}-3k\omega a
^{6\omega-2}+\frac{P_T}{12}(3\omega-1)a^{6\omega}\rho-
\frac{2\Lambda}{3}a^{6\omega}~.
\end{equation}
To obtain above two equations of motion we have used the
continuity equation $\dot{\rho}+\frac{\dot{a}}{a}\rho(1+\omega)=0$
as well as the result of its integration i.e. $\rho=\rho_{0}a^{-3
(1+\omega)}$ (we have set $\rho_{0}=1$) due to perfect fluid energy
density. As the final words in this
section, there are two points for more attention: firstly, the terms including
numerical constant $C$ in the rhs of the Friedmann and Raychaudhuri
equations, address the matter contribution of the SB scalar field.
Secondly, these generalized dynamical equations are not the same things as
we know from the standard cosmology. More precisely, the absence of a scalar
potential term $V(\phi)$ in the action (\ref{e2-1}) distinguishes equations
(\ref{e2-18}) and (\ref{e2-19}) from their standard counterparts which are
derived from original (standard) inflationary action including a scalar field minimally
coupled to Einstein gravity with self-interaction potential
term, see Refs.~\cite{Ellis:2004, Barrow:2003}.

\section{First order dynamic system analysis of ESU
without cosmological constant, $\Lambda=0$ }
Now we perform a first order
dynamical system analysis on ESU(s) emerging from the cosmological model at hand.
We consider the case with $\Lambda=0$. It is expected that at the end of the analysis
there will be a clear answer to the question \emph{``is it possible a well realization
of the EC paradigm to circumvent the Big-Bang
singularity within the mentioned toy model
universe? If yes, then under what conditions?''}.
In the first step, using the background equations
(\ref{e2-18}) and (\ref{e2-19}) we extract
the relevant static solutions (ESU(s)) and check the
validity of physical conditions for each of these solutions.
As the next step, we investigate the stability of derived
static solutions by employing a first order dynamical system
analysis. At the final step, we discuss on the issue of
\emph{"natural transition from ESU to the standard cosmology evolution"}.\\

From equations (\ref{e2-18}) and (\ref{e2-19}), one can show that the static solution
(with $a=a_{s}$ and $\dot{a}=0=\ddot{a}$) fulfills the following equations for the case
with $\Lambda=0$
\begin{equation}\label{e3-1}
\frac{C}{6}a_{s}^{6\omega-6}-ka_{s}^{6\omega-2}
+\frac{P_T}{6}a_{s}^{6\omega}\rho_{s}=0~,
\end{equation}
and
\begin{equation}\label{e3-2}
\frac{C}{6}a_{s}^{6\omega-6}-3k\omega a_{s}^{6\omega-2}
+\frac{P_T}{12}(3\omega-1)a_{s}^{6\omega}\rho_{s}=0~,
\end{equation}  respectively.
We begin our dynamical system analysis for the case of flat model, $k=0$.
Solving equation (\ref{e3-1}) we arrive at the following relation
\begin{equation}\label{e3-3}
\rho_{s}=-\frac{C}{P_{T}a_{s}^{6}}~,
\end{equation} between $a_s$ and $\rho_s$.
Here, to save the positivity of the energy density
($\rho_{s}>0$), we have to choose only the negative values
for $P_T$. There is no fixed value for the critical point
$a_s$ and it can take any positive value. Having the relation (\ref{e3-3}),
one finds that equation (\ref{e3-2}) for the case of flat geometry
satisfies only if $\omega=1$\footnote{EoS as $\rho=p$ points out an exotic
component known as stiff fluid which for the first time was studied
by Zeldovich \cite{Zeldovich:1962}. There are a large number of studies on
cosmological models with stiff fluid.
For instance, in some inhomogeneous cosmological models with EoS parameter $\omega=1$
some exact solutions without singularity have been extracted \cite{Mars:1995,
Fernandez:1997, Fernandez:2002}. In Ref. \cite{Titus:2014} some other important cosmological implications of the stiff
fluid are presented.}.
In order to investigate stability of physical critical
points (among them is the ESU), our strategy, inspired by
\cite{Reza:2009, Wu:2010, Kaituo:2014, Huang:2015}, is to reduce the dynamics of the cosmological model at hand to
the first order by the following change of variables
\begin{equation}\label{e3-4}
x_1=a,~~~~~~x_2=\dot{a}~.
\end{equation}
In this situation, we have
\begin{eqnarray}\label{e3-5}
\begin{array}{ll}
\dot{x}_1=x_2=Y_1(x_1,x_2)~~~\mbox{and}~~~
\dot{x}_2=\frac{C}{6}(3\omega-2)x_{1}^{6\omega-5}-
3k\omega x_{1}^{6\omega-1}+\frac{P_T}{12}(3\omega-1)
x_{1}^{6\omega+1}\rho=Y_2(x_1,x_2)~.
\end{array}
\end{eqnarray}
To treat the stability of ESU(s), one just needs to determine eigenvalues $\lambda^2$
of the following Jacobian matrix
\begin{eqnarray}\label{e3-6}
J\bigg(Y_1(x_1,x_2),Y_2(x_1,x_2)\bigg)=\left(
 \begin{array}{cc}
 \frac{\partial Y_1}{\partial x_1} & \frac{\partial Y_1}{\partial x_2}\\
 \frac{\partial Y_2}{\partial x_1} & \frac{\partial Y_2}{\partial x_2}\\
 \end{array}
  \right)~~\Rightarrow~~~\lambda^2=\frac{\partial Y_2}{\partial x_1}|_{(a_s,\rho_s)}
  \end{eqnarray}
where only in the case with $\lambda^2<0$ one can assign a stable ESU to the
relevant physical critical point\footnote{Physically, $\lambda^2<0$
means that the critical point $(a_s, 0)$ is a stable concentric center
so that if one applies a small perturbation, that point
does not collapse. However, $\lambda^2>0$ can address an unstable saddle point.
Note however that $\lambda^2>0$ does not necessarily means that it is
an unstable saddle point always. For instance if in Jacobian matrix (\ref{e3-6}),
$\frac{\partial Y_1}{\partial x_1} \neq\frac{\partial Y_2}{\partial x_2}$
and $\frac{\partial Y_1}{\partial x_2}=0=\frac{\partial Y_2}{\partial x_2}$,
then $\lambda^2>0$ addresses a stable node.}
\cite{Huang:2015}. In a spatially flat universe case, $k=0$, we
have
\begin{equation}\label{e3-7}
\lambda^2=\frac{C}{6}(3\omega-2)(6\omega-5)a_{s}^{6\omega-6}+
\frac{P_T}{12}(3\omega-1)(6\omega+1)a_s^{6\omega}\rho_s~,
\end{equation}
where after application of the relation (\ref{e3-3}) as well as the
condition $\omega=1$ into (\ref{e3-7}), we find $\lambda^2<0$.
This means that if the ESU is perturbed from its original
condition by a small perturbation, then the universe experiences an
infinite series of oscillations around the the initial state, since
by fixing $\omega=1$ in (\ref{e2-19}), we have
 \begin{equation}\label{e3-7*}
\ddot{a}-\Big(\frac{C+P_T}{6}\Big)a=0~.
\end{equation}
Since $P_T<0$, then if $|P_{T}|>C$, this equation
results in an oscillatory solution as $e^{i\sqrt{|\frac{C+P_T}
{6}|}t}$. Although this result seems to be interesting, it
suffers from lack of transition to the inflation epoch.
As has been mentioned previously, the ESU only for the EoS parameter
$\omega=1$ is physically meaningful. Therefore, one expects it never
get out of the stable situation to entre into the inflation epoch as
the universe evolves. In another words, despite the realization of a
stable ESU in this model with a spatially flat geometry, the initial
singularity yet remains.\\
We continue our analysis by focusing on the contribution of the non-flat
spatial geometries ($k\neq0$). From the background equations (\ref{e3-1})
and (\ref{e3-2}) in the presence of the curvature constant we obtain the
following critical points
\begin{equation}\label{e3-8}
a_{s}^4=\frac{C}{2k}\Big(\frac{1-\omega}{1+3\omega}
\Big) ~~~\mbox{and}~~~~ \rho_s=\frac{1}{P_{T}}\Big
(\frac{6k}{a_{s}^{2}}
-\frac{C}{a_{s}^{6}}
\Big)~.
\end{equation}
With a straightforward calculation one can find that
in particular for the case with $k=+1$, both of the above critical
points are physically meaningful (i.e. $a_{s}>0$ and
$\rho_{s}>0$) provided that
\begin{eqnarray}\label{e3-9}
\begin{array}{ll}
P_{T}>0,~~~-\frac{1}{3}<\omega<\frac{1}{3}~,\\
P_{T}<0,~~~\frac{1}{3}\leq\omega<1~.
\end{array}
\end{eqnarray}
Now, the relevant eigenvalues $\lambda^2$ can be written as
\begin{eqnarray}\label{e3-10}
\lambda^2=\frac{C}{6}(3\omega-2)(6\omega-5)\Big(\frac{C(1-\omega)}{2+6\omega}
\Big)^{\frac{3\omega}{2}-\frac{3}{2}}-3\omega(6\omega-1)\Big(\frac{C(1-\omega)}
{(2+6\omega)}\Big)^{\frac{3\omega}{2}-\frac{1}{2}} +\nonumber\\
\frac{C^{\frac{3\omega}{2}-\frac{1}{2}}}{12}(3\omega-1)(6\omega+1)\Big\{6\big
(\frac{1-\omega}
{2+6\omega}\big)^{\frac{3\omega}{2}-\frac{1}{2}}-
\big(\frac{1-\omega}{2+6\omega}\big)^{\frac{3\omega}{2}-\frac{3}{2}}\Big\}~.
\end{eqnarray}
By regarding the first constraint mentioned in (\ref{e3-9}),
one infers that here there is no possibility of extracting
stable ESU since for any desired value $C>0$, we have $\lambda^2>0$.
\begin{figure}
\begin{center}
\begin{tabular}{c}\hspace{-1cm}\epsfig{figure=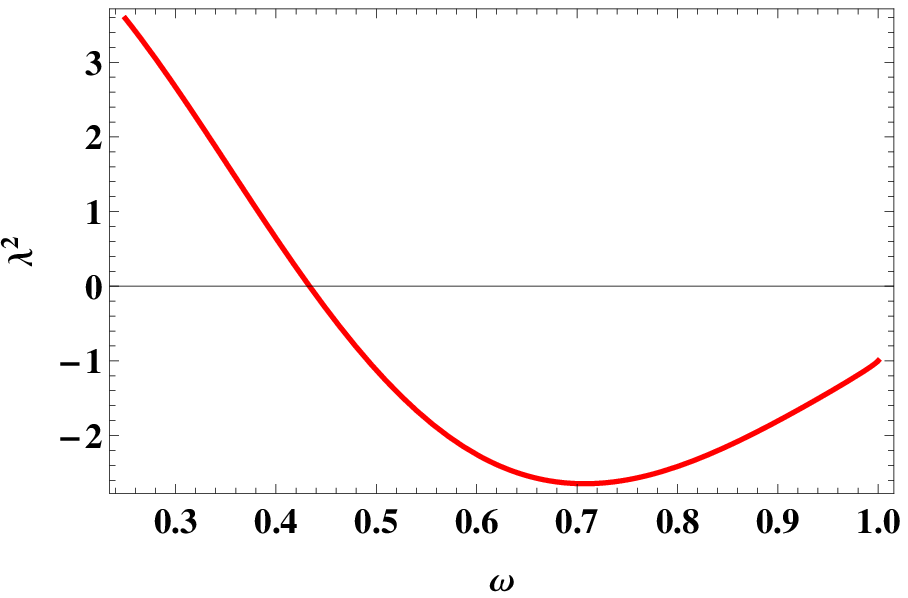,width=6.5cm}
\hspace{1cm} \epsfig{figure=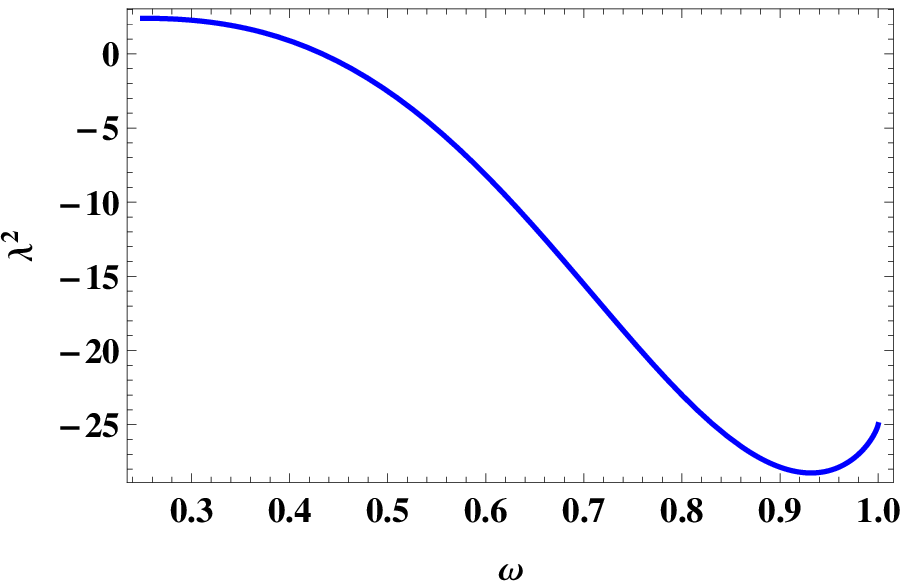,width=6.5cm}
\end{tabular}
\end{center}
\caption{\footnotesize {\emph{ $\lambda^2$ versus $\omega$
for the case with $k=+1$ and by satisfying the seconded constraint in
(\ref{e3-9}). We have set the numerical value as $C=1$ and $C=25$ for
the left and right panels respectively. Note that the general behavior of
$\lambda^2-\omega$ in the case of fixing the numerical values $0<C<25$ is similar to the left panel
while if $C\geq25$ it is similar to the right panel.} }}
\label{fig:1}
\end{figure}
However, the second constraint in (\ref{e3-9}) addresses a
different situation, see Fig. 1. We see that by choosing some
numerical values for $C$, ESU may experience both stable and
unstable phases with time evolution of the universe. More precisely,
for any values limited to the interval $0<C<25$, there are the possibility of having
both positive and negative values for $\lambda^2$ in terms of the allowed values of
$\omega$. However, for $C\geq25$, the eigenvalue is utterly negative.
As an example,
if we fix the value of $C$ as $C=1$ (the left panel of Fig. 1)
in the toy model at hand, then ESU is stable if $0.4<\omega<1$.
While with growth of the cosmic time and subsequently reducing
the EoS parameter to $\omega=1/3$, the state changes towards
instability. This changing phase of the critical point (from
stability to instability) with cosmic time can be thought as a
graceful transition of the ESU to the standard cosmological evolution
as is expected and required. On the other hand, if we fix the value of
$C$ as $C=25$ (the right panel of Fig. 1) and even larger values, the
universe remains in a stable static phase forever without beginning the
standard thermal history. This situation clearly is shown in Fig. 2 for
three cases $\omega=1/3,~2/3,~4/5$. Now, if the scale factor $a_s$
be affected by a slight perturbation, there will be no decay and the
universe  oscillates around an equilibrium point. Note that here the Big
Bang singularity problem still endures.\\

From equation (\ref{e3-8}), for case with negative spatial curvature
$k=-1$, the critical points $\rho_s$ and $a_s$ are physical only in case
with
\begin{eqnarray}\label{e3-11}
P_{T}<0,~~~\omega<-\frac{1}{3}~.
\end{eqnarray}
As a result, the eigenvalues $\lambda^2$ take the following form
\begin{eqnarray}\label{e3-12}
\lambda^2=\frac{C}{6}(3\omega-2)(6\omega-5)\Big(\frac{C(\omega-1)}{2+6\omega}
\Big)^{\frac{3\omega}{2}-\frac{3}{2}}+3\omega(6\omega-1)\Big(\frac{C(\omega-1)}
{(2+6\omega)}\Big)^{\frac{3\omega}{2}-\frac{1}{2}} -\nonumber\\
\frac{C^{\frac{3\omega}{2}-\frac{1}{2}}}{12}(3\omega-1)(6\omega+1)\Big\{6\big
(\frac{\omega-1}
{2+6\omega}\big)^{\frac{3\omega}{2}-\frac{1}{2}}+
\big(\frac{\omega-1}{2+6\omega}\big)^{\frac{3\omega}{2}-\frac{3}{2}}\Big\}\,.
\end{eqnarray}
One can simply show that for any values of $C>0$ and $P_{T}<0$,
by fixing the values of the EoS parameter as $\omega<-\frac{1}{3}$, the eigenvalue
$\lambda^2$ is positive. Therefore, from the cosmological model at hand with $k<0$,
in the absence of cosmological constant $\Lambda$, a sustainable ESU cannot be extracted.
In Table 1 we have listed all the possibilities in this cosmological setup with $\Lambda=0$
which results in a steady ESU.\\
\begin{figure}
\begin{center}
\begin{tabular}{c}\hspace{-1cm}\epsfig{figure=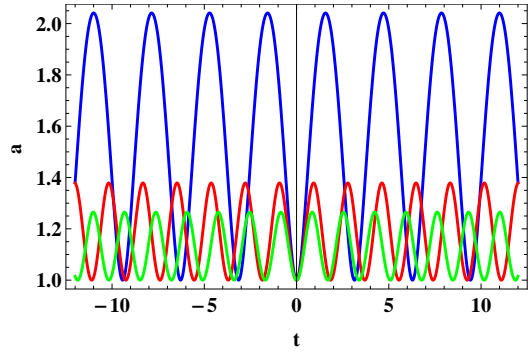,width=7cm}
\end{tabular}
\end{center}
\caption{\footnotesize {\emph{The scale factor, $a$, versus the cosmic time $t$
for the case with $k=+1$ corresponding to three EoS parameter
$\omega=1/3$ (blue curve), $\omega=2/3$ (red curve)
and $\omega=4/5$ (green curve)
with numerical values $C=25$ and $P_{T}=-1$.} }}
\label{fig:2}
\end{figure}
\begin{table}
\caption{\label{Tab1} \emph{Summary of the stable physical critical points
(ESU) for $\Lambda=0$. The case with $k=+1$ is the case that leads to fulfilling
the conditions of the EC in order to eliminate the Big-Bang
singularity. }}
\begin{center}
 \begin{tabular}{|r|r|r|}
  \hline
  & Physical Critical Points & Stability Conditions\\
  \hline
  $k=0$ & $\rho_{s}=-\frac{C}{P_{T}a_{s}^{6}}$,~~~$a_{s}$ can be any positive
  desired value &  $P_T<0,~~\omega=1$ \\
  \hline
  $k=+1$ & $a_{s}^4=\frac{C}{2}\Big(\frac{1-\omega}{1+3\omega}
\Big) ~~~\mbox{and}~~~~ \rho_s=\frac{1}{P_{T}}\Big(\frac{6}{a_{s}^{2}}-\frac{C}{a_{s}^{6}}
\Big)$&  $P_{T}<0,~~~\frac{1}{3}\leq\omega<1$  \\
  \hline
  \end{tabular}
\end{center}
   \end{table}
\\

\section{First order dynamical system analysis of the ESU with a cosmological
constant}
Regarding the contribution of the cosmological constant $\Lambda$ in
the matter part of the action (\ref{e2-1}), by the same way as the previous
section here we focus further on the dynamics of ESUs. Static solutions in
the presence of a cosmological constant, $\Lambda$, should satisfy the following
equations
\begin{equation}\label{e4-1}
\frac{C}{6}a_{s}^{6\omega-6}-ka_{s}^{6\omega-2}
+\frac{P_T}{6}a_{s}^{6\omega}\rho_{s}-\frac{\Lambda}{6} a_{s}
^{6\omega}=0~,
\end{equation}
and
\begin{equation}\label{e4-2}
\frac{C}{6}a_{s}^{6\omega-6}-3k\omega a_{s}^{6\omega-2}
+\frac{P_T}{12}(3\omega-1)a_{s}^{6\omega}\rho_{s}-\frac{2\Lambda}
{3}a_{s}^{6\omega}=0~.
\end{equation}
As before, let us begin our analysis from the case
of $k=0$ which leads to the critical points as
\begin{equation}\label{e4-3}
a_{s}^{-6}=\frac{\Lambda}{C}\Big(\frac
{9-3\omega}{1-3\omega}\Big) ~~~~\mbox{and}~~~~
\rho_s=-\frac{\Lambda}
{P_T}\Big(\frac{8}{1-3\omega}\Big)~.
\end{equation}
These critical points have physical meaning if
\begin{eqnarray}\label{e4-4}
\begin{array}{ll}
\Lambda>0,~~~ P_{T}<0,~~~\omega<\frac{1}{3}~,\\
\Lambda<0,~~~ P_{T}<0,~~~\frac{1}{3}<\omega<3~,\\
\Lambda>0,~~~ P_{T}>0,~~~\omega>3~.
\end{array}
\end{eqnarray}
Adding the contribution of $\Lambda$ to the first order
differential equation (\ref{e3-6}), the eigenvalue $\lambda^2$
generally can be read as
\begin{equation}\label{e4-5}
\lambda^2=\frac{C}{6}(3\omega-2)(6\omega-5)x_{1}^{6\omega-6}
-3k\omega(6\omega-1)x_{1}^{6\omega-2}+\frac{P_T}{12}(3\omega-1)
(6\omega+1)x_{1}^{6\omega}\rho-\frac{2\Lambda}{3}(6\omega+1)x_{1}
^{6\omega}\,.
\end{equation}
Putting the critical points (\ref{e4-3}) into (\ref{e4-5}),
for the case of $k=0$, the final form of the eigenvalue $\lambda^2$
is obtained as
\begin{equation}\label{e4-6}
\lambda^2=\frac{C}{6}(3\omega-2)(6\omega-5)\Big(\frac
{\Lambda(9-3\omega)}{C(1-3\omega)}\Big)^{1-\omega}~.
\end{equation}
Note that with three constraints listed in (\ref{e4-4}), the
contribution of $\Lambda$ can not lead to the
derivation of a sustained static solution (ESU) since $\lambda^2>0$. \\

In what follows, we consider non-flat cases with $k\neq0$.
The critical point corresponding to energy density
$\rho_{s}$ reads as
\begin{equation}\label{e4-7}
\rho_s=\frac{1}{P_T}\Big(\Lambda+\frac{6k}{a_{s}^{2}}-
\frac{C}{a_{s}^{6}}\Big)~.
\end{equation}
However, in deriving $a_{s}^{-2}$ we dealing with a
cubic equation as follows
\begin{equation}\label{e4-8}
\frac{C}{4}(1-\omega)z^3-\frac{3k}{2}(\omega+\frac{1}{3})
z+\frac{\Lambda}{4}(\omega-3)=0~,
\end{equation} in which $ z\equiv a_{s}^{-2}$.
Generally, the solutions of the above equation can be
given via the following expression
\begin{equation}\label{e4-9}
\Delta=-4\alpha \beta^{3}-27\alpha^{2}\gamma^{2}~,
\end{equation}where $\alpha=\frac{C}{4}(1-\omega),
~\beta=-\frac{3k}{2}(\omega+\frac{1}{3})$ and $\gamma
=\frac{\Lambda}{4}(\omega-3)$.
In particular, by setting $k=\pm1$, the above expression
gives $\Delta<0$, if
\begin{eqnarray}\label{e4-10}
\begin{array}{ll}
\Lambda>0~~~\mbox{or}~~~ \Lambda<0,~~~~\omega<-1,~~~~-1\leq\omega
<-\frac{1}{3},~~~~\omega>1,\\
\Lambda>\sqrt{\frac{-32-288\omega-864(\omega^2+\omega^3)}{-243C+405C\omega-189
C\omega^2+27C\omega^3}}~~~\mbox{or}~~~ \Lambda<-\sqrt{\frac{-32-288\omega-864
(\omega^2+\omega^3)}{-243C+405C\omega-189
C\omega^2+27C\omega^3}},~ -\frac{1}{3}<\omega<1~;
\end{array}
\end{eqnarray}
and
\begin{eqnarray}\label{e4-10*}
\begin{array}{ll}
\Lambda>0~~~\mbox{or}~~~ \Lambda<0,~~~~-\frac{1}{3}<\omega<1,\\
\Lambda>\sqrt{\frac{32+288\omega+864(\omega^2+\omega^3)}{-243C+405C\omega-189
C\omega^2+27C\omega^3}}~~~\mbox{or}~~~ \Lambda<-\sqrt{\frac{32+288\omega+864
(\omega^2+\omega^3)}{-243C+405C\omega-189C\omega^2+27C\omega^3}},~\omega<-
\frac{1}{3},~1<\omega<3~.
\end{array}
\end{eqnarray} corresponding to the closed and open spatial geometries,
respectively. The negative sign of $\Delta$ means that
the cubic equation (\ref{e4-8}) has only a real solution (two other
solutions are imaginary and unacceptable) which is given as
\begin{equation}\label{e4-11}
\frac{1}{a_{s}^{2}}=-\frac{1}{3\alpha}\Big(D+\frac{\Delta_{0\pm}}{D}\Big)~,
\end{equation}
so that
\begin{equation}\label{e4-12}
\Delta_{0\pm}=\pm\frac{9C}{8}(1-\omega)(\omega+\frac{1}{3}),~~~\Delta_1=
\frac{27}{64} C^{2}\Lambda(\omega-3)(1-\omega)^{2},~~~\mbox{and}~~~D=\Big(\frac{\Delta_1
+\sqrt{\Delta_{1}^{2}-4
\Delta_{0\pm}^{3}}}{2}\Big)^{\frac{1}{3}}~.
\end{equation}
Notice that the sign $\pm$ in these relations
corresponds to the positive and negative spatial
curvature respectively. Now let us to see whether
these critical points satisfy the physical conditions
($\frac{1}{a_{s}^{2}}>0$ and $\rho_s>0$) or not.
We then infer that the above mentioned critical
points represent a factual ESU with respect
to the following conditions
\begin{eqnarray}\label{e4-13}
\begin{array}{ll}
\Lambda>0,~~~ P_T>0,~~~~\omega<-1,~~~~-1\leq\omega<-\frac{1}{3},\\
\Lambda<0,~~~ P_T<0,~~~~\omega>1~,
\end{array}
\end{eqnarray}
and
\begin{eqnarray}\label{e4-13*}
\begin{array}{ll}
\Lambda>0,~~~ P_T<0,~~~~-\frac{1}{3}<\omega<1,\\
\end{array}
\end{eqnarray} corresponding to the cases $k=\pm1$ respectively.
As is seen from these conditions, the second case listed in (\ref{e4-10})
and (\ref{e4-10*}) can not result in a factual ESU. As a result inserting
the above critical points into (\ref{e4-5}), we get
\begin{eqnarray}\label{e4-14}
\lambda_{\pm}^2=\frac{C}{6}(3\omega-2)(6\omega-5)\Big(-\frac{1}
{3\alpha}\big(D+\frac{\Delta_{0\pm}}{D}\big)\Big)^{3-3\omega}
\mp3\omega(6\omega-1)\Big(-\frac{1}{3\alpha}\big(D+\frac
{\Delta_{0\pm}}{D}\big)\Big)
^{1-3\omega} -\hspace{2cm}\nonumber\\
\Big(-\frac{1}{3\alpha}\Big(D+\frac{\Delta_{0\pm}}{D}\Big)\Big)
^{-3\omega}\Big\{\frac{-2\Lambda}{3}(6\omega+1)+
\frac{(3\omega-1)(6\omega+1)}{12}\Big(\Lambda-\frac{2}{\alpha}
\big(D+\frac{\Delta_{0\pm}}{D}\big)+
\frac{C}{27\alpha^3}\big(D+\frac{\Delta_{0\pm}}{D}\big)^3\Big)\Big\}\,.
\end{eqnarray}
Interestingly, we find that for $k=1$, the sign of $\lambda^2$ in the case
where the conditions listed in (\ref{e4-13}) are satisfied, depends on the
values of $C$ and $\Lambda$. In other words, while by attributing some values
for $C$ and $\Lambda$, we get a stable ESU with $\lambda_{+}^2<0$, setting some
other values can lead to an unstable ESU, i.e. $\lambda_{+}^2>0$. Moreover,
we note that under the conditions (\ref{e4-13}), no transition occurs from stability
to instability (as the EoS parameter $\omega$ reduces, the sign of $\lambda_{+}^2$
does not change to positive one) to enter the standard thermal history of the universe.
However, for the case of $k=-1$, if the conditions (\ref{e4-13*}) are satisfied,
we are dealing with an unstable ESU ($\lambda_{-}^{2}>0$), so that this
result is independent of fixing what amount to the $C$ and $\Lambda$. \\

The relation (\ref{e4-9}) with $\omega=1$ gives the conditions for fulfillment of
$\Delta=0$. In this case, we are dealing with the critical points as
\begin{equation}\label{e4-15}
\frac{1}{a_{s}^{2}}=-\frac{\Lambda}{4k}~~~~~\mbox{and}~~~~
\rho_s=\frac{1}{P_T}\Big(-\frac{\Lambda}{2}-\frac{C\Lambda^3}
{64k^3}\Big)~.
\end{equation}
It is not difficult to demonstrate that the above mentioned critical points
are physical provided that
\begin{eqnarray}\label{e4-16}
\begin{array}{ll}
k=1,~~~~\Lambda<0,~~~~ P_T>0,\\
k=-1,~~~~\Lambda>\sqrt{\frac{32}{C}},~~~~P_T>0,\\
k=-1,~~~~0<\Lambda<\sqrt{\frac{32}{C}},~~~~P_T<0~.
\end{array}
\end{eqnarray}
Now by having the critical points (\ref{e4-15}) and by
setting $\omega=1$ in (\ref{e4-5}) we arrive at the following
expression
\begin{equation}\label{e4-17}
\lambda^2=\frac{13C}{6}+\frac{784k^3}{3\Lambda^3}-480
\sqrt{-\frac{k^{7}}{\Lambda^5}}~,
\end{equation} for the eigenvalue $\lambda^2$. Surprisingly,
for all three physical conditions listed in (\ref{e4-16}) we
face with a sustainable ESU i.e. $\lambda^2<0$. So, as is
displayed in Fig. 3, in response to external small perturbations,
ESU does not destroy since it fluctuates steadily. This means that
in this situation also despite the existence of a stable static solution,
there is no possibility to exit from ESU and consequently the early singularity
problem stays unresolved.
\begin{figure}
\begin{center}
\begin{tabular}{c}\hspace{-1cm}\epsfig{figure=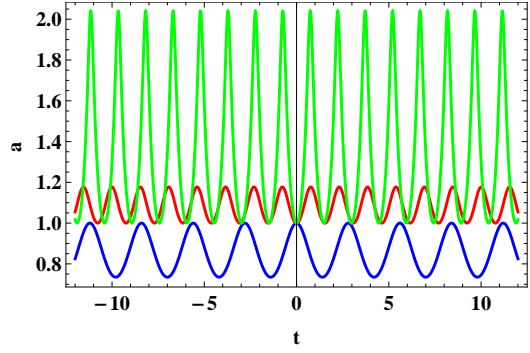,width=7cm}
\end{tabular}
\end{center}
\caption{\footnotesize {\emph{ The scale factor versus the cosmic
time in the presence of $\Lambda$ and with the EoS parameter $\omega=1$
corresponding to three physical conditions (\ref{e4-16}) from the blue
to green curve respectively. We have set the numerical values $C=8,~P_{T}=1,~\Lambda=-1$
(blue curve), $C=8,~P_{T}=1,~\Lambda=5$ (red curve)
and $C=8,~P_{T}=-1,~\Lambda=3/2$ (green curve).} }}
\label{fig:2}
\end{figure}
As the final case, provided that
\begin{eqnarray}\label{e4-17}
\begin{array}{ll}
k=1,~~~~-\frac{1}{3}<\omega<1,~~~~\Lambda>0~~~~\mbox{or}~~~~\Lambda<0,\\
k=-1,~~~~~\omega<-1,~~~~-1\leq\omega<-\frac{1}{3},~~~~\Lambda>0~~~~
\mbox{or}~~~~\Lambda<0~,
\end{array}
\end{eqnarray} we have $\Delta>0$ which addresses the existence of
three separate real solutions for the cubic equation (\ref{e4-8}).
These three real solutions include
\begin{eqnarray}\label{e4-18}
\begin{array}{ll}
(Z_+)_1=A_+\cos(\frac{\phi_+}{3}),~~~
(Z_+)_2=A_+\cos(\frac{\phi_+}{3}+\frac{2\pi}{3}),~~~
(Z_+)_3=A_+\cos(\frac{\phi_+}{3}+\frac{4\pi}{3})~,
\end{array}
\end{eqnarray}
and
\begin{eqnarray}\label{e4-19}
\begin{array}{ll}
(Z_-)_1=A_-\cos(\frac{\phi_-}{3}),~~~
(Z_-)_2=A_-\cos(\frac{\phi_-}{3}+\frac{2\pi}{3}),~~~
(Z_-)_3=A_-\cos(\frac{\phi_-}{3}+\frac{4\pi}{3})~,
\end{array}
\end{eqnarray} for the closed and open universes
respectively where
\begin{equation}\label{e4-20}
A_{\pm}=\sqrt{\pm\frac{4}{3C}\frac{6\omega+2}{1-\omega}}~,~~~
\phi_{\pm}=\cos^{-1}\Big(\frac{\frac{3\Lambda(3-\omega)}{C(1-\omega)}}
{\sqrt{\pm\frac{4}{3}(\frac{6\omega+2}{C(1-\omega)})^3}}\Big)~.
\end{equation}
Among the three solutions presented in (\ref{e4-18}) and (\ref{e4-19}),
only the first and third ones satisfy the physical
condition $\frac{1}{a_{s}^{2}}>0$ with the following conditions
\begin{eqnarray}\label{e4-19*}
\begin{array}{ll}
(Z_+)_1>0,~~~\mbox{if}~~~\Lambda>0,~~~\mbox{or}~~~\Lambda<0~~~\mbox{and}~~~|\Lambda|\ll1\\
(Z_+)_3>0,~~~\mbox{if}~~~\Lambda<0~~~\mbox{and}~~~|\Lambda|\ll1\\
(Z_-)_1>0,~~~\mbox{if}~~~\Lambda>0,~~~\mbox{or}~~~\Lambda<0~~~\mbox{and}~~~|\Lambda|\ll1\\
(Z_-)_3>0,~~~\mbox{if}~~~\Lambda>0~~~\mbox{and}~~~|\Lambda|\ll1~.
\end{array}
\end{eqnarray}
Now by substituting the mentioned physical critical points
into (\ref{e4-8}), one gets
\begin{eqnarray}\label{e4-20*}
\begin{array}{ll}
(\rho_{s+})_1=\frac{1}{P_T}\Big(\Lambda+6A_+\cos(\frac{\phi_+}{3})-
CA_{+}^{3}\cos^{3}(\frac{\phi_+}{3})\Big)~,\\
(\rho_{s+})_3=\frac{1}{P_T}\Big(\Lambda+6A_+\cos(\frac{\phi_+}{3}+\frac{4\pi}{3})-
CA_{+}^{3}\cos^{3}(\frac{\phi_+}{3}+\frac{4\pi}{3})\Big)~,\\
(\rho_{s-})_1=\frac{1}{P_T}\Big(\Lambda-6A_-\cos(\frac{\phi_-}{3})-
CA_{-}^{3}\cos^{3}(\frac{\phi_-}{3})\Big)~,\\
(\rho_{s-})_3=\frac{1}{P_T}\Big(\Lambda-6A_-
\cos(\frac{\phi_-}{3}+\frac{4\pi}{3})-
CA_{-}^{3}\cos^{3}(\frac{\phi_-}{3}+\frac{4\pi}{3})\Big)~.
\end{array}
\end{eqnarray}
By focusing on the first two
cases relevant to $k=+1$, we find that for $(\rho_{s+})_1$ and
$(\rho_{s+})_3$, the positive energy condition holds if in addition to
fulfillment of the relevant constraints (\ref{e4-19*}), we set $P_T<0$
and $P_T>0$ in the cosmological model at hand, respectively. Eventually,
the relevant expressions of the eigenvalues $\lambda^2$ take the following
forms
\begin{eqnarray}\label{e4-21}
(\lambda_{+}^2)_{1}=\frac{C}{6}(3\omega-2)(6\omega-5)\Big(A_+\cos(\frac{\phi_+}{3})
\Big)^{3-3\omega}-3\omega(6\omega-1)\Big(A_+\cos(\frac{\phi_+}{3})
\Big)^{\frac{1}{2}-3\omega} -\hspace{2cm}\nonumber\\
\Big(A_+\cos(\frac{\phi_+}{3})\Big)^{-3\omega}\Big\{\frac{-2\Lambda}{3}(6\omega+1)+
\frac{(3\omega-1)(6\omega+1)}{12}\Big(\Lambda+6A_+\cos(\frac{\phi_+}{3})
-C\big(A_+\cos(\frac{\phi_+}{3})\big)^3\Big)\Big\}~,
\end{eqnarray}
and
\begin{eqnarray}\label{e4-22}
(\lambda_{+}^2)_{3}&=&\frac{C}{6}(3\omega-2)(6\omega-5)\Big(A_+\cos(\frac{\phi_+}{3}
+\frac{4\pi}{3})\Big)^{3-3\omega}
-3\omega(6\omega-1)\Big(A_+\cos(\frac{\phi_+}{3}+\frac{4\pi}{3})
\Big)^{\frac{1}{2}-3\omega} -\hspace{2cm}\nonumber\\ &&
\Big(A_+\cos(\frac{\phi_+}{3}+\frac{4\pi}{3})\Big)^{-3\omega}
\left\{\frac{-2\Lambda}{3}(6\omega+1)+\frac{(3\omega-1)(6\omega+1)}{12}
\Big(\Lambda+6A_+\cos(\frac{\phi_+}{3}+\frac{4\pi}{3})\right. \nonumber\\ & &\left.
-C\big(A_+\cos(\frac{\phi_+}{3}+\frac{4\pi}{3})\big)^3\Big)\right\}~.
\end{eqnarray}
respectively. Equation (\ref{e4-21}) tells us that regarding the first constraint in
(\ref{e4-19*}), a complete realization of emergent cosmology can be possible. As an
example see Fig. 4 (the left panel). The physical critical point $(Z_+)_1$ is an
interesting static solution with amazing feature that it can join to the standard
cosmological evolution in a fascinating manner. However, as can be seen from the
right panel of Fig. 4, the critical point $(Z_+)_3$ by the satisfaction of the
second constraint in (\ref{e4-19*}), represents a permanent static phase in
pre-inflation epoch without joining the expected standard scenario.
\begin{figure}
\begin{center}
\begin{tabular}{c}\hspace{-1cm}\epsfig{figure=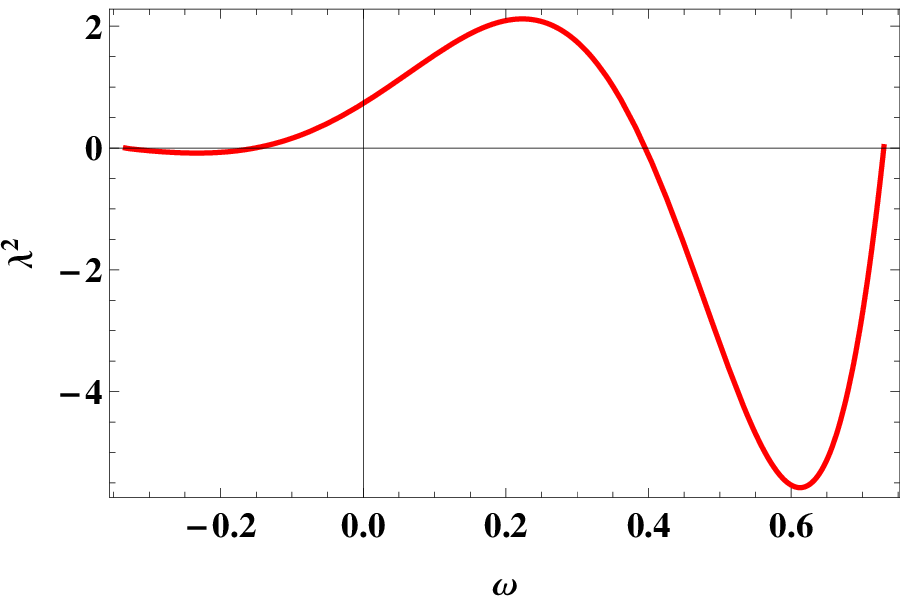,width=7cm}
\hspace{1cm} \epsfig{figure=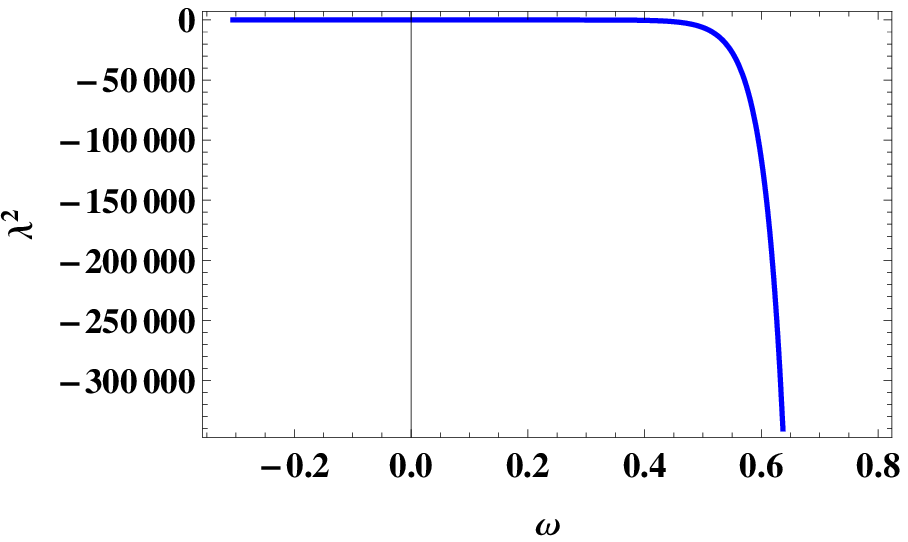,width=7cm}
\end{tabular}
\end{center}
\caption{\footnotesize {\emph{The plot of $\lambda^2$ versus $\omega$ in a closed
universe model ($k=+1$) corresponding to two physical static solutions
$ (Z_+)_1$ (left panel) and $ (Z_+)_3$ (right panel) by setting the numerical values
$C=50,~\Lambda=0.0001$.} }}
\label{fig:1}
\end{figure}
By concerning on the open geometry ($k=-1$), we stress that $(\rho_{s-})_1$
and $(\rho_{s-})_3$ can satisfy the positive energy condition if besides
fulfilling the relevant constraints (\ref{e4-19*}), we set $P_T<0$ in
our cosmological model. As a result, here we present the final form of the
expressions for $\lambda^2$ in this case
\begin{eqnarray}\label{e4-23}
(\lambda_{-}^2)_{1}=\frac{C}{6}(3\omega-2)(6\omega-5)\Big(A_-\cos(\frac{\phi_-}{3})
\Big)^{3-3\omega}+3\omega(6\omega-1)\Big(A_-\cos(\frac{\phi_-}{3})
\Big)^{\frac{1}{2}-3\omega} -\hspace{2cm}\nonumber\\
\Big(A_-\cos(\frac{\phi_-}{3})\Big)^{-3\omega}\Big\{\frac{-2\Lambda}{3}(6\omega+1)+
\frac{(3\omega-1)(6\omega+1)}{12}\Big(\Lambda-6A_-\cos(\frac{\phi_-}{3})
-C\big(A_-\cos(\frac{\phi_+}{3})\big)^3\Big)\Big\}
\end{eqnarray}
and
\begin{eqnarray}\label{e4-24}
(\lambda_{-}^2)_{3}&=&\frac{C}{6}(3\omega-2)(6\omega-5)\Big(A_-\cos(\frac{\phi_-}{3}
+\frac{4\pi}{3})\Big)^{3-3\omega}
+3\omega(6\omega-1)\Big(A_-\cos(\frac{\phi_-}{3}+\frac{4\pi}{3})
\Big)^{\frac{1}{2}-3\omega} -\hspace{2cm}\nonumber\\ &&
\Big(A_-\cos(\frac{\phi_-}{3}+\frac{4\pi}{3})\Big)^{-3\omega}
\left\{\frac{-2\Lambda}{3}(6\omega+1)+\frac{(3\omega-1)(6\omega+1)}{12}
\Big(\Lambda-6A_-\cos(\frac{\phi_-}{3}+\frac{4\pi}{3})\right. \nonumber\\ & &\left.
-C\big(A_-\cos(\frac{\phi_-}{3}+\frac{4\pi}{3})\big)^3\Big)\right\}\,,
\end{eqnarray}
\begin{table}
\caption{\label{Tab1} \emph{Summary of the stable physical critical points
(ESU) in the presence of $\Lambda$. The case marked with * is the case that
leads to fulfilling the conditions of the EC in order to eliminate the Big-Bang
singularity.}}
\begin{center}
 \begin{tabular}{|r|r|r|}
  \hline
  &  Physical critical points&Stability conditions\\
  \hline
    $k=+1$ & $a_{s}^{-2}=-\frac{1}{3\alpha}\Big(D+\frac{\Delta_0}{D}\Big),~~~\rho_s
    =\frac{1}{P_T}\Big(\Lambda+\frac{6k}{a_{s}^{2}}-
\frac{C}{a_{s}^{6}}\Big)$ &  $\begin{array}{ll}
\Lambda>0,~P_T>0,~\omega<-1,~-1\leq\omega<-\frac{1}{3}\\
\Lambda<0,~~ P_T<0,~~\omega>1~
\end{array}$  \\ \hline
  & $a_{s}^{-2}=-\frac{\Lambda}{4},~\rho_s=\frac{1}{P_T}
\Big(-\frac{\Lambda}{2}-\frac{C\Lambda^3}{64}\Big)~$ &  $\omega=1,~~\Lambda<0,~ P_T>0$  \\ \hline
*& ~$(Z_+)_1,~~~(\rho_{s+})_1$ & $-\frac{1}{3}<\omega<1,~\Lambda>0~\mbox{or}~\Lambda<0,~|\Lambda|\ll1,~P_T<0$ \\ \hline
& $(Z_+)_3,~~~(\rho_{s+})_3$ & $-\frac{1}{3}<\omega<1,~\Lambda<0,~|\Lambda|\ll1,~P_T>0$ \\ \hline
  $k=-1$ & $a_{s}^{-2}=\frac{\Lambda}{4},~~~\rho_s=\frac{1}{P_T}
\Big(-\frac{\Lambda}{2}+\frac{C\Lambda^3}{64}\Big)~$ &  $\begin{array}{ll}
\omega=1,~~~\Lambda>\sqrt{\frac{32}{C}},~~~~P_T>0,\\
\omega=1,~~~0<\Lambda<\sqrt{\frac{32}{C}},~~~~P_T<0 ~
\end{array}$ \\
 \hline
  \end{tabular}
\end{center}
   \end{table}
respectively. Unlike the case of positive spatial curvature, here both
physical critical points $(Z_-)_1$ and $(Z_-)_3$ lead to $\lambda_{-}^2>0$.
In Table 2, we have listed all the possible cases leading to a stable ESU in
the presence of a cosmological constant $\Lambda$.

In comparison between Tables 1 and 2 we see that adding the cosmological
constant term to the matter sector of the action make more and severe
requirements for stability of ESU since now we are encountering a wider
parameter space of the model. In each of these two Tables only one case
has the potential for well realization of EC paradigm within underlying
model universe. In this case the Big-Bang singularity is completely
removed. This is the main achievement of this detailed study.

\section{Global dynamical system analysis of the model}
So far we have performed a local dynamical system analysis of the ES solutions.
We saw that in the framework of a local dynamical system analysis there is
the possibility of removing the early universe singularity through a full realization
of the EC scenario. However, for the EC paradigm to be viable one definitely needs a
supplementary analysis which covers the whole phase space i.e. a global dynamical system analysis.
Hence, to complete the study of the dynamical system analysis of the model we focus on the ES solutions at infinite phase space. It is done
through the following coordinate transformations
\begin{eqnarray}\label{esub-1}
(x_1,x_2)\Rightarrow(X_1,X_2):~X_1=\frac{1}{x_1}=\frac{1}{a},~~X_2=\frac{x_2}{x_1}=\frac{\dot{a}}{a};~0<X_1<+\infty
,~-\infty<
X_2<+\infty
\end{eqnarray} in the phase plane to cover a circle at infinity. Now the background equations (\ref{e3-1})
and (\ref{e3-2}) are re-expressed as follows
\begin{equation}\label{esub-2}
\frac{C}{6}X_{1s}^{6-6\omega}-kX_{1s}^{2-6\omega}
+\frac{P_T\rho_{s}}{6}X_{1s}^{-6\omega}=0~,
\end{equation}
and
\begin{equation}\label{esub-3}
\frac{C}{6}X_{1s}^{6-6\omega}-3k\omega X_{1s}^{2-6\omega}
+\frac{P_T\rho_{s}}{12}(3\omega-1)X_{1s}^{-6\omega}=0~,
\end{equation}  respectively. As a result, the autonomous
system transforms to\footnote{Note that for this autonomous
system in a finite domain of the phase space there are also
other ES solutions which previously have been taken into account.}
\begin{eqnarray}\label{esub-4}
\dot{X_1}=-X_1X_2,~~~~\dot{X_2}=\frac{C}{6}(3\omega-1)X_1^{6-6\omega}-3k\omega X_1^{2-6\omega}+
\frac{p_T}{12}\rho(3\omega-1)X_1^{-6\omega}-X_2^{2}~.
\end{eqnarray}
The conditions for realization of the ES solution ($ a=a_s,~ \rho=\rho_s,~\dot{a}=0$) require $\big(X_{1s}, X_{2s}=0\big)$
as the critical points on a circle at infinity for the autonomous system (\ref{esub-4}).
In this global analysis with the phase space with infinite domain, the existence of
a stable global de Sitter (dS) attractor is essential for the EC paradigm to be successful.
Therefore, in the following analysis we fix $\omega=-1$.\\
In the case of $k=0$, by setting $X_{1s}=\big(-\frac{p_T\rho_s}{C}\big)^{1/6}$ (with $p_T<0$ and $\rho_s>0$)
into the relevant Jacobian matrix, the squared eigenvalue is obtained to be a positive value which addresses
an unstable global dS solution. As a result, the relevant fixed point cannot play the role of a dS-attractor. \\
For the cases $k=\pm1$, we get the following real critical points
\begin{eqnarray}\label{esub-5}
X_{1s\pm}=\sqrt{\mp 2^{4/3} \bigg(C^2 p_T \rho_s\big(1 + \sqrt{1\pm\frac{32}{C p_T^2 \rho_s^2}}\bigg)\bigg)^{-1/3}
+\bigg(\frac{ p_T\rho_s}{2C} \bigg(1 + \sqrt{1\pm \frac{32}{c p_T^2 \rho_s^2}}\bigg)\bigg)^{1/3}}
\end{eqnarray}
on a circle at infinity, respectively. We found that for these two fixed points to represent global dS-attractors (i.e. $\lambda^2<0$), the
integration constant $C$ should be negative, which is in contrary to what the equation (\ref{e2-16}) imposes on the model.
Therefore, in these cases existence of a global dS-attractor to have a successful realization of EC scenario is not possible.
\begin{figure}
\begin{center}
\begin{tabular}{c}\hspace{-1cm}\epsfig{figure=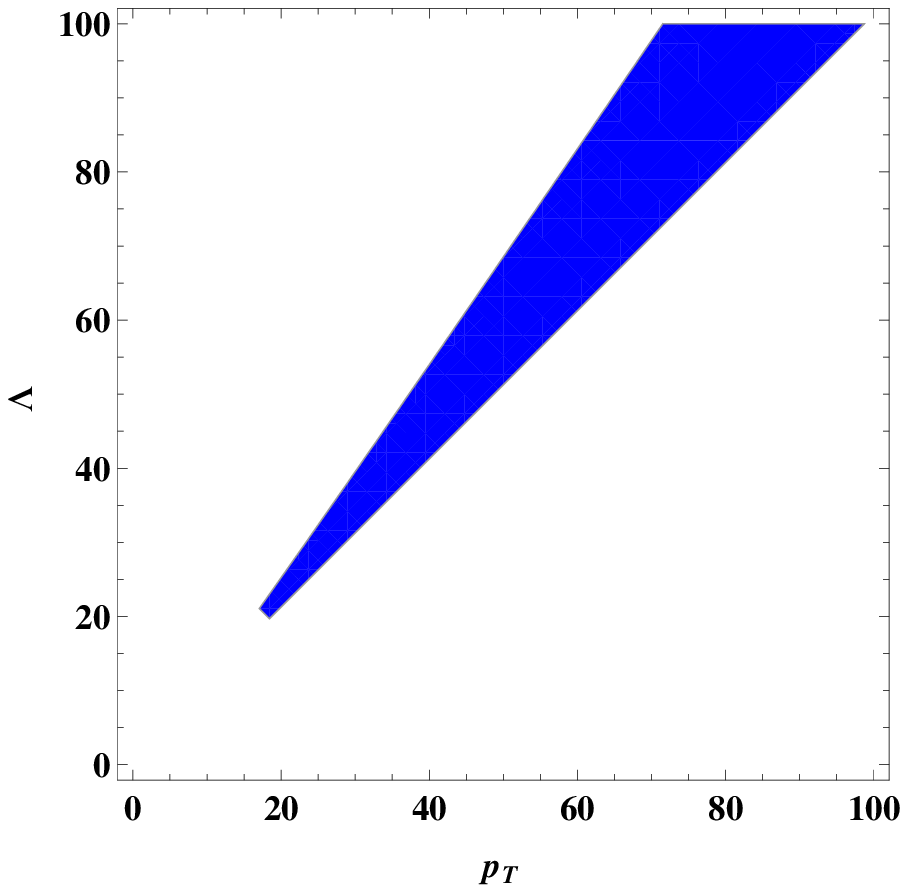,width=7cm}
\hspace{1cm} \epsfig{figure=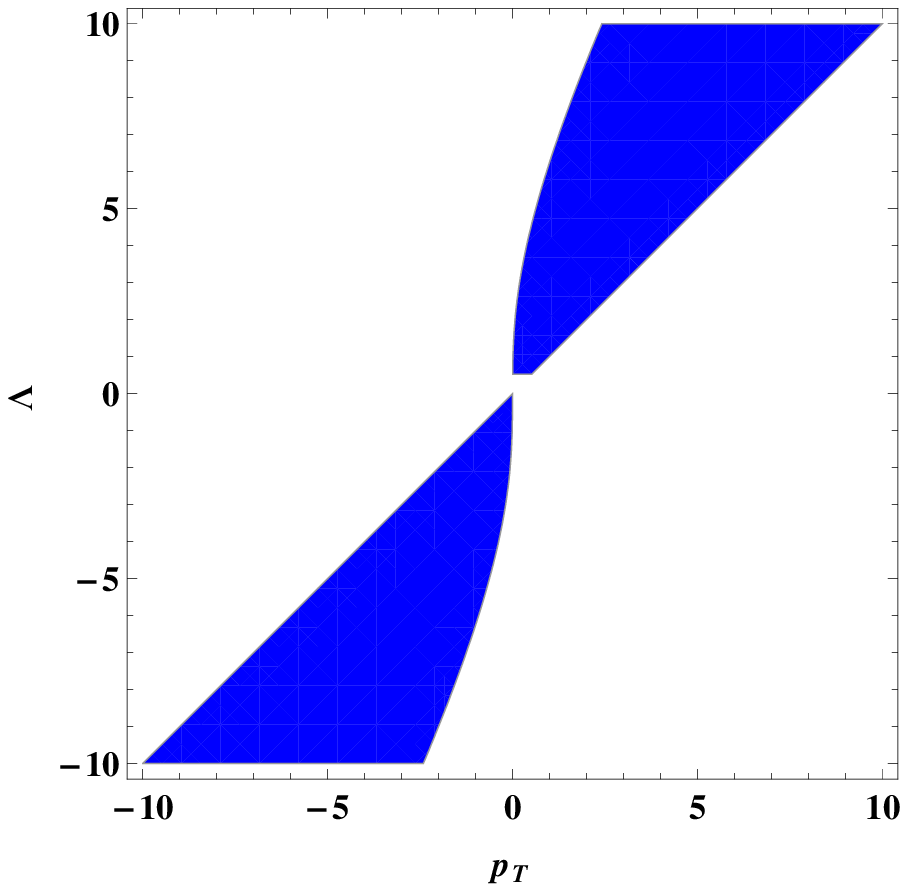,width=7cm}
\end{tabular}
\end{center}
\caption{\footnotesize {\emph{Existence regions for dS-attractor of critical points (\ref{esub-8}) relevant to a spatially closed
(left panel) and spatially open (right panel) universes in the $(p_T-\Lambda)$ parameter space. Here we have set positive values for $C$ and $\rho_s$.} }}
\label{fig:1}
\end{figure}
To proceed further, we continue our global stability analysis now with a new ingredient: the presence of a cosmological constant $\Lambda$.
With the same coordinate transformations, this time the autonomous system transforms to
\begin{eqnarray}\label{esub-6}
\dot{X_1}=-X_1X_2,~~~~\dot{X_2}=-\frac{2C}{3}X_1^{12}+3k X_1^{8}-
(\frac{p_T}{3}\rho+\frac{2\Lambda}{3})X_1^{6}-X_2^{2}~.
\end{eqnarray}
The squared eigenvalue of the above autonomous system reads as
\begin{eqnarray}\label{esub-7}
\lambda^2=8CX_{1s}^{12}-24kX_{1s}^8+(2p_T\rho_s-4\Lambda)X_{1s}^6\,.
\end{eqnarray}
By setting the fixed points $X_{1s}=(\frac{4\Lambda}{C})^{1/6},~~
\rho_s=-\frac{2\Lambda}{p_T}$ for the case of flat spatial geometry,
we find $\lambda^2=\frac{96\Lambda^2}{C}>0$. So, we obtain the real critical points
\begin{eqnarray}\label{esub-8}
X_{1s\pm}=\sqrt{\mp 2^{4/3}\bigg(9C^2(\Lambda- p_T \rho_s)\bigg(1 + \sqrt{1\pm\frac{864}{C (\Lambda-p_T \rho_s)^2}}\bigg)\bigg)^{-1/3}
+\bigg(\frac{ (\Lambda-p_T\rho_s)^2}{2C} \bigg(1 + \sqrt{1\pm \frac{864}{C (\Lambda-p_T \rho_s)^2}}\bigg)\bigg)^{1/3}}
\end{eqnarray} on a circle at infinity for the non-flat cases $k=\pm1$, respectively. As can be seen from Fig. 5
this time, unlike the previous cases, by tuning some values of $p_T$ and $\Lambda$, fixed points (\ref{esub-8}) could represent
global dS-attractors. Therefore, to have a successful EC scenario in a perfect fluid scalar-metric scenario, the existence of a cosmological constant $\Lambda$
is necessary.

\section{EC scenario from effective potential approach}

Following our classical stability analysis, here, inspired by \cite{Jcap:2014} we intend to see the
ESU stability issue and subsequently a full realization of the
EC scenario from another approach known as \emph{``effective potential
formalism''}. Indeed, by viewing ESU as a classical particle which is affected
by a one-dimensional potential as $U$, we shall derive the required information by
concerning on the stability status of the ESU. \\

For this purpose, the first Friedmann equation can be
rewritten in terms of the effective potential $U(a)$
as follows
\begin{eqnarray}\label{e5-1}
\frac{1}{2}\dot{a}^2+U(a)=E\equiv0~,~~~~~\mbox{where}~~~~U(a)=-\frac{C}
{12}a^{6\omega-4}+\frac{k}{2}a^{6\omega}
-\frac{P_T}{12}a^{3\omega-1}+\frac{\Lambda}{12}a^{6\omega+2}
\end{eqnarray}
Translating the conditions required for realization of the ESU (i.e. $\dot{a}=0=\ddot{a}$)
in the context of effective potential formalism leads us to the following conditions
 \begin{eqnarray}\label{e5-2}
U(a_s)=0~~~~\mbox{and}~~~U'(a_s)=0~,
\end{eqnarray}
which can be read as follows
\begin{eqnarray}\label{e5-3}
U(a_s)=-\frac{C}{12}a_s^{6\omega-4}+\frac{k}{2}a_s^{6\omega}
-\frac{P_T}{12}a_s^{3\omega-1}+\frac{\Lambda}{12}a_s^{6\omega+2}=0~~,
\end{eqnarray}
and
\begin{eqnarray}\label{e5-4}
U'(a_s)=-\frac{C}{6}(3\omega-2)a_s^{6\omega-5}+3k\omega a_s^{6\omega-1}
-\frac{P_T}{12}(3\omega-1)a_s^{3\omega-2}+\frac{\Lambda}{6}(3\omega+1)
a_s^{6\omega+1}=0
\end{eqnarray} respectively.
At the first sight, one can see that under any condition $a_s=0$ can be a
trivial extremum point (EP) for the above equations. It is not so hard to prove
that the aforementioned EP cannot be a physical point.
However, equations (\ref{e5-3}) and (\ref{e5-4}) under some constraints on the
free parameters $C$ and $P_T$ suggest another EP as $a_s=1$. In what follows,
once in the absence and then in the presence of cosmological constant $\Lambda$,
for the underlying cosmological toy model we investigate possible realization
of a non-singular EC paradigm by concerning on the stability/instability
status of the local EP that represents an ESU.\\

\subsection{$\Lambda=0$}
By neglecting the contribution of the cosmological constant $\Lambda$ in our toy model,
for the case of flat spatial geometry $k=0$, there is an EP as $a_s=1$ moreover an unacceptable
case $a_s=0$ if $C=-P_T$ and $w=1$. However, it is clear that with the mentioned
constraints we have $U''(a_s=1)=0$ which means that $a_s=1$ is not a local EP, rather is
a saddle point. As a result, here the stability status of the saddle point $a_s=1$ is complicated and unclear.\\

However, for the non-flat cases ($k=\pm1$), the EP $a_s=1$ can be saved provided that
\begin{eqnarray}\label{e5-5}
P_T=\pm\frac{8}{1-\omega}~~~~~\mbox{and}~~~~~ C=\pm\frac{2+6\omega}{\omega-1}~.
\end{eqnarray}
These conditions are very valuable in the sense that they make our analysis no
longer dependent on the model free parameters ($C$ and $P_T$) unlike the phase analysis
presented in previous sections. By applying aforementioned conditions, we come to
the following equations
\begin{eqnarray}\label{e5-6}
U''(a_s)=-\frac{2 a_s^{-3 + 3 \omega} (-2 + 3 \omega) (-1 + 3 \omega)}{3 (1 - \omega)}
\pm\frac{a_s^{-6 + 6 \omega} (-5 + 6 \omega) (-4 + 6 w) (2 + 6 \omega)}{12 (-1 + \omega)}
\pm 3 a_s^{-2 + 6 \omega} \omega (-1 + 6 \omega)
\end{eqnarray} which are corresponding to the closed (+ sign) and open (- sign)
spatial geometries, respectively. The stability/instability status of the
ESU ($a_s=1$) is recognizable through $U''(a_s=1)>0$ and $U''(a_s=1)<0$, respectively.
Therefore, we arrive at the following constraints on parameter $\omega$
\begin{eqnarray}\label{e5-7}
w>-\frac{1}{3}~~~~\Rightarrow ~~~~U''(a_s=1)>0\\ \nonumber
w<-\frac{1}{3}~~~~\Rightarrow ~~~~U''(a_s=1)<0
\end{eqnarray}
and
\begin{eqnarray}\label{e5-8}
w<-\frac{1}{3}~~~~\Rightarrow ~~~~U''(a_s=1)>0\\ \nonumber
w>-\frac{1}{3}~~~~\Rightarrow ~~~~U''(a_s=1)<0
\end{eqnarray} for the cases of $k=\pm1$, respectively.
The above constraints suggest that only for the case
with $k=+1$ it is possible to exit the initial static mode
in order to enter into the inflation period.  As is shown
qualitatively in Fig. 6, by decreasing the EoS parameter
$\omega$ (with this idea in mind that in infinitely past
time i.e. $t\rightarrow-\infty$ it was a constant value)
from $\omega>-1/3$ towards $\omega<-1/3$ as universe moves
forward, the stable EP $a_s=1$ converts into an unstable
counterpart so that through an explicit violation of SEC it
prepares to enter the accelerating non-singular EC.
We note that within the context of
effective potential formalism, the unstable state suggests
two different paths for ESU: because of small inhomogeneities
ESU moves towards standard expanding thermal history
i.e. EC or contracts to an initial singularity, as we can see
qualitatively in the right panel of Fig. 6.\\
\begin{figure}
\begin{center}
\begin{tabular}{c}\hspace{-1cm}\epsfig{figure=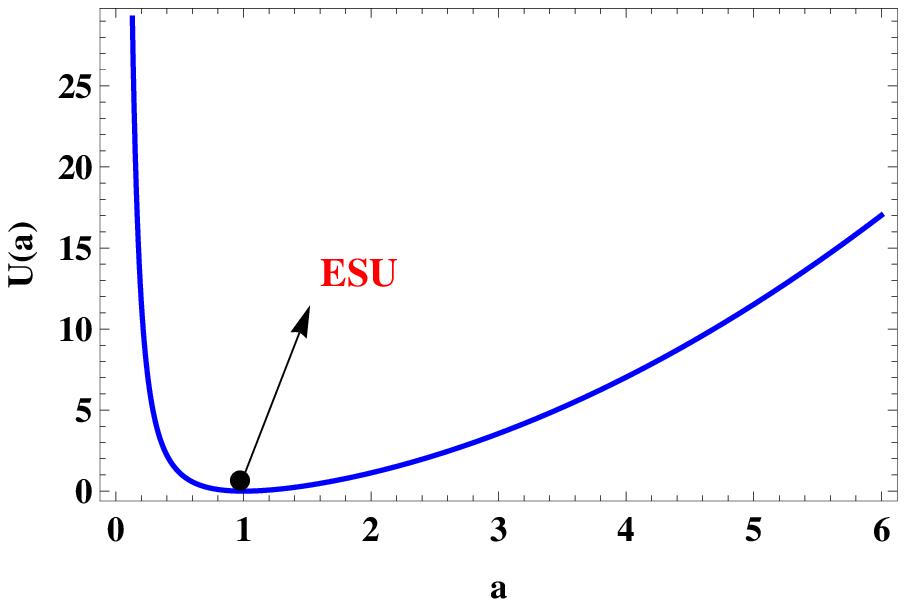,width=6.5cm}
\hspace{1cm} \epsfig{figure=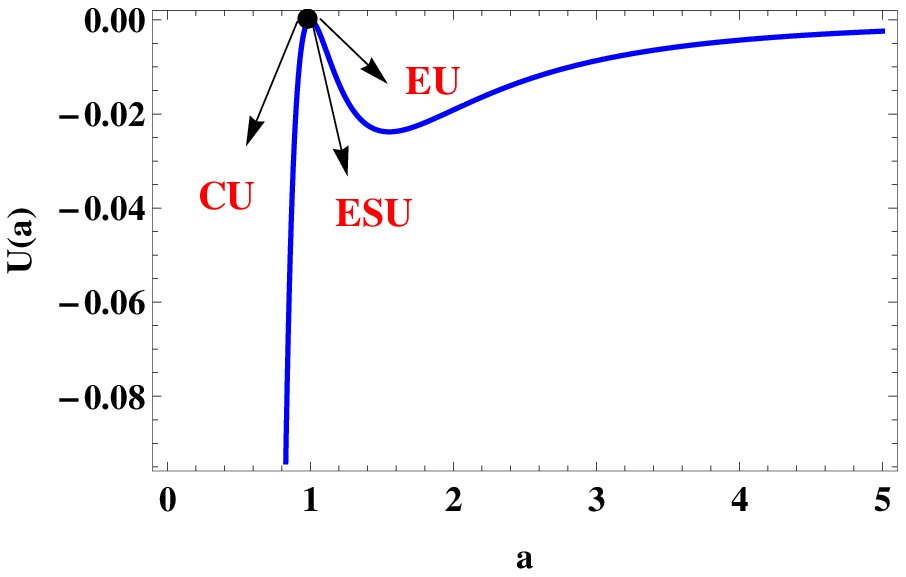,width=6.5cm}
\end{tabular}
\end{center}
\caption{\footnotesize \emph{The scheme of the effective potential for a closed universe raised
by the toy model at hand with $\Lambda=0$ consisting of two types matter fields with $w=\frac{1}{3}$
(left panel) and $w=-\frac{2}{3}$ (right panel). The minimum of the effective potential in the left
panel represents a stable ESU while its counterpart's maximum in the right panel corresponds to an unstable
ESU since against an small perturbations it moves either into an accelerating emergent universe (EU)
or into a contracting universe (CU).}}
\label{fig:E}
\end{figure}

\subsection{ $\Lambda\neq0$}
By adding the cosmological constant $\Lambda$ into our cosmological
model, the EP $a_s=1$ can be represents as ESU for the case $k=0$ if
\begin{eqnarray}\label{e5-9}
C=\Lambda\big(\frac{\omega+1}{\omega-1}\big),~~~~~~~~P_T=\frac{2\Lambda}{1-\omega}
\end{eqnarray}
By putting this expressions into the effective potential we get
 \begin{eqnarray}\label{e5-10}
U''(a_s)=\frac{ a_s^{-6 +6 \omega} (1 + \omega) (-5 + 6 \omega) (-4 + 6\omega)\Lambda+a_s^{-3
+3 \omega} (-2 + 3 \omega) (-1 + 3 w) (8 + 2\Lambda)}{12 ( \omega-1)}
\end{eqnarray}
In the $\omega-\Lambda$ plane the EP $a_s=1$ shows a stable or unstable
ESU (see the two figures in upper panel of Fig. 6 with
relevant descriptions).
\begin{figure}[ht]
\begin{tabular}{c}
\epsfig{figure=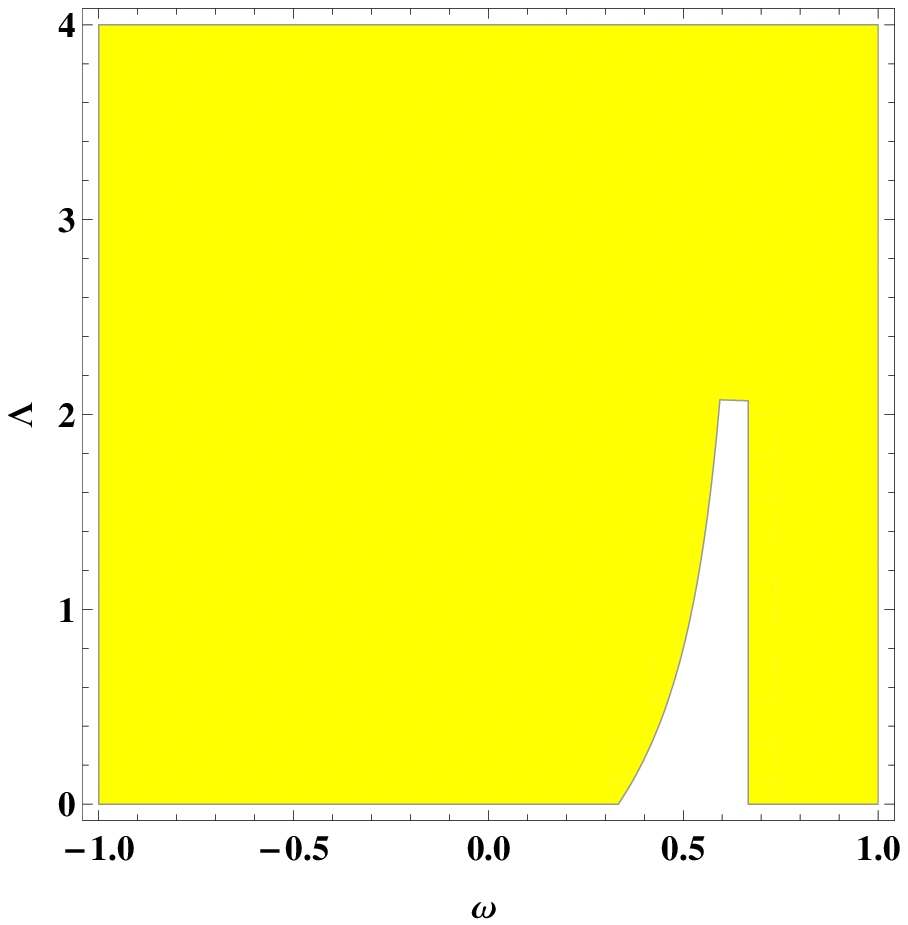,width=5.7cm}
\epsfig{figure=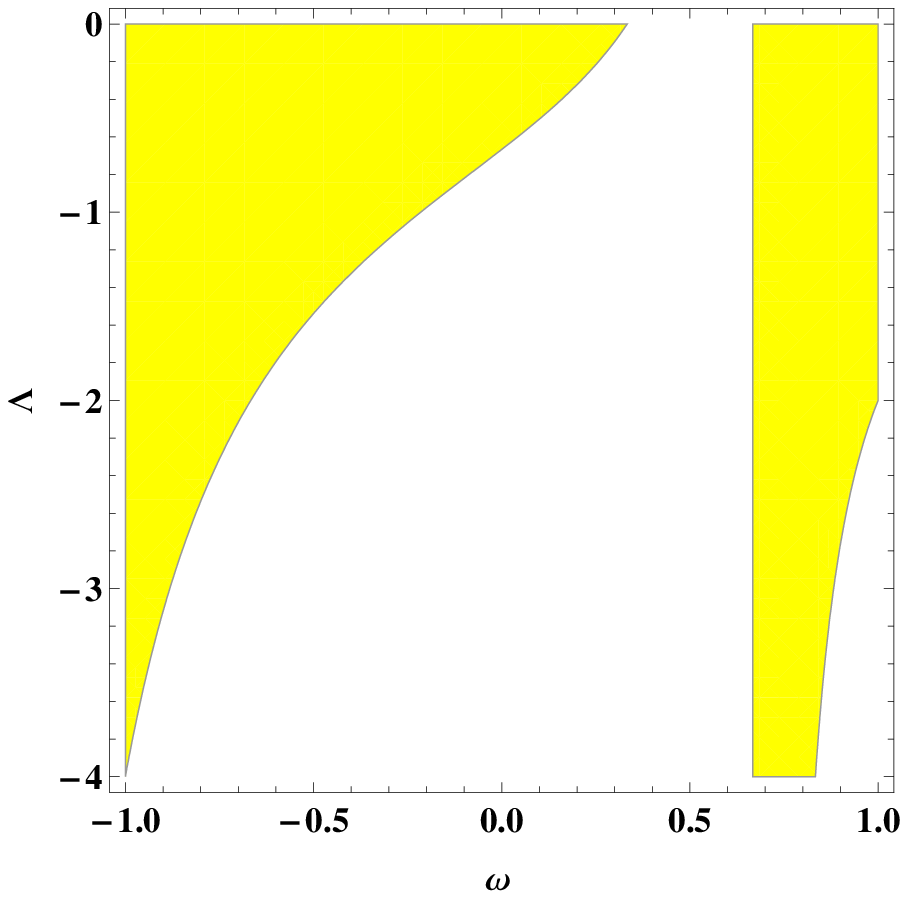,width=5.7cm}\\
\epsfig{figure=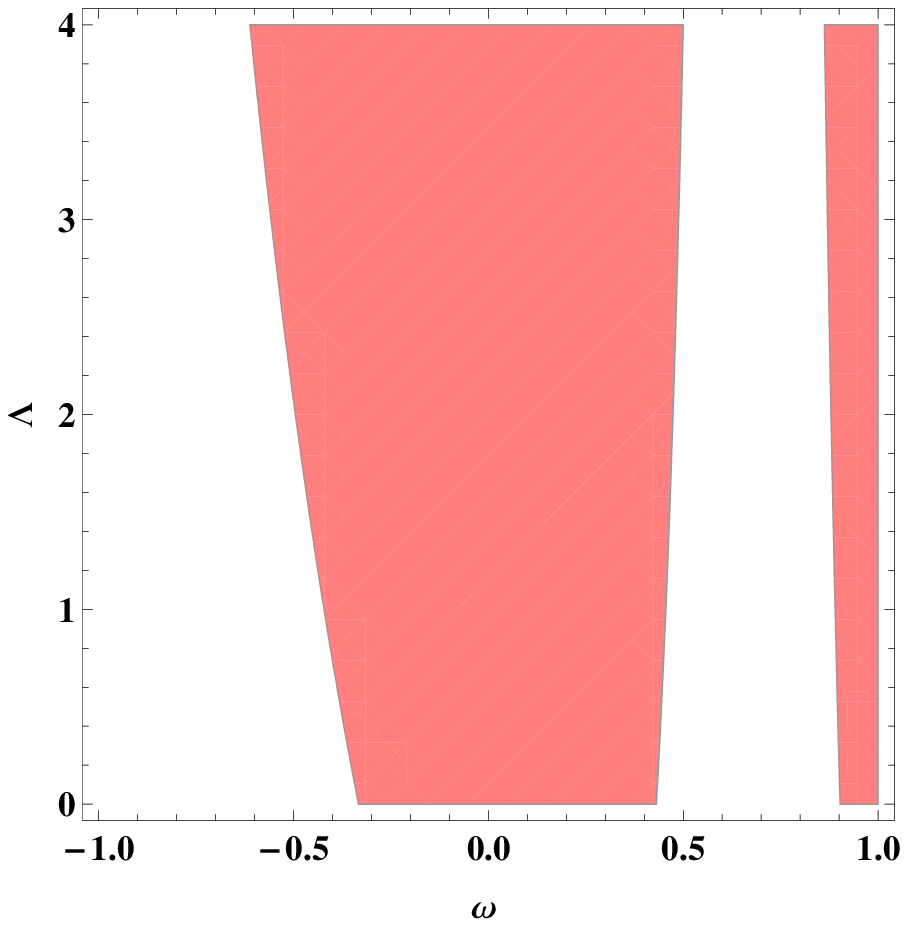,width=5.7cm}
\epsfig{figure=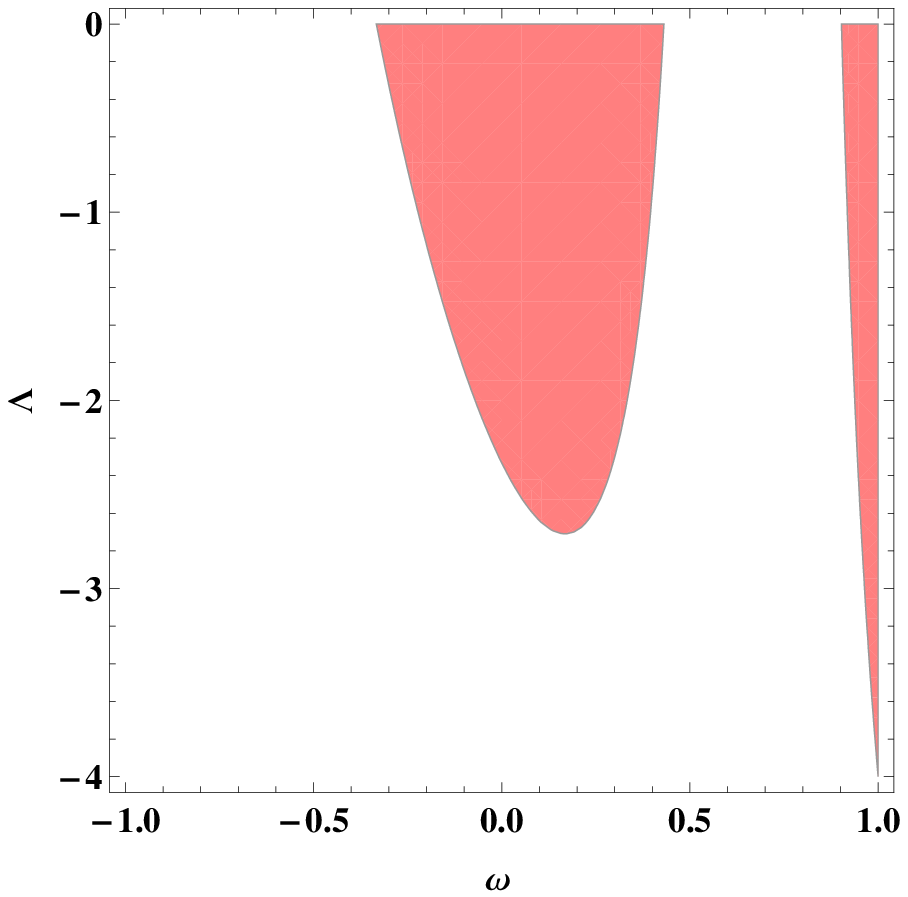,width=5.7cm}\\
\epsfig{figure=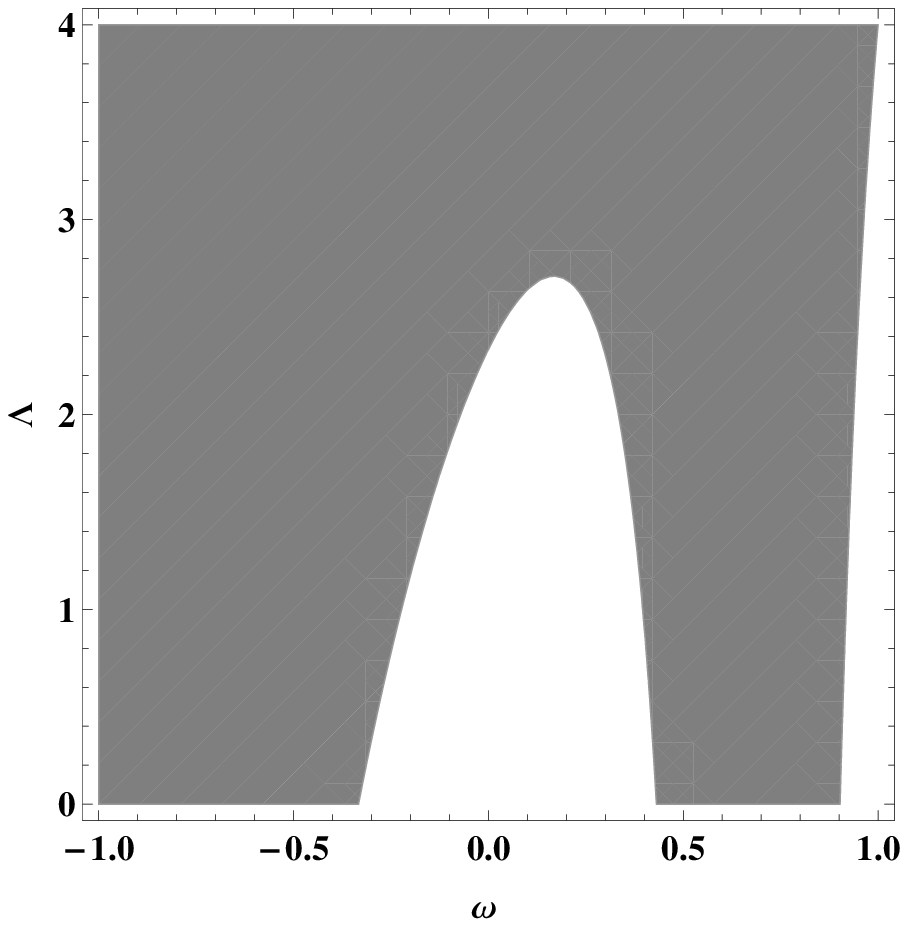,width=5.7cm}
\epsfig{figure=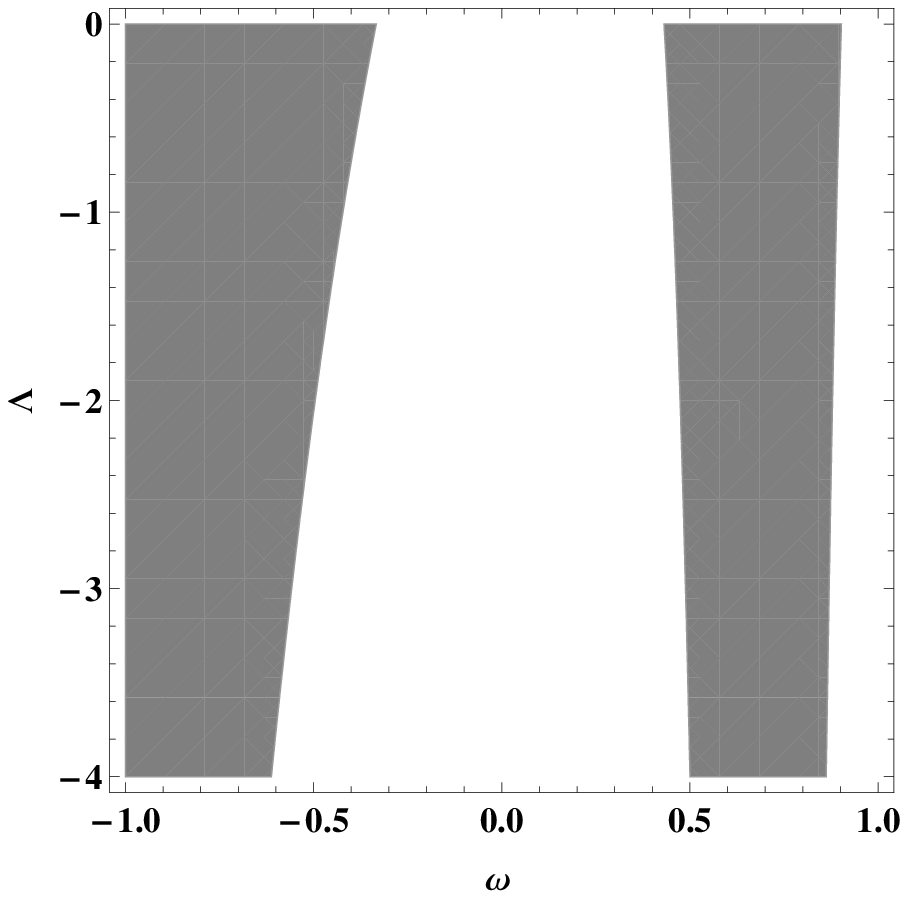,width=5.7cm}
\end{tabular}
\caption{\footnotesize \emph{Existence and Stability/instability regions
for critical point $a_s=1$ relevant to a spatially flat (the upper couple), closed (the middle couple)
and open (the lower couple) universe in the $(\omega-\Lambda)$ parameter space.
In each couple, the left and right panels are devoted to the positive and negative values
of the cosmological constant $\Lambda$ respectively. Colorless and colored areas represent
regions in which ESU is stable and unstable respectively.} }
\label{fig:E}
\end{figure}

As is shown qualitatively in the upper couple of the figures in Fig. 7, in the
presence of a positive or negative cosmological constant for a flat geometry universe,
a graceful realization of the EC is possible. Focusing on the same argument as already
been raised, we see in both plots that by reduction of $\omega$ (as the universe evolves)
the EP $a_s=1$ as a representative of the ESU, transits from stability region to the instability
region to joint to the standard cosmology (here we mean inflation period). Note that for the
case with $\Lambda>0$, it is possible to have stability/instability switching even without
violating SEC. If $\Lambda<0$, only for
small values of $\Lambda$ this possibility may be present so that for larger
values, SEC explicitly violates. \\

Using the conditions (\ref{e5-3}) and (\ref{e5-4}) for the cases of
$k=\pm1$, we can rewrite the free parameters $C$ and $P_T$ in terms
of $\omega$ and $\Lambda$ as follows
\begin{eqnarray}\label{e5-11a}
C=\frac{\Lambda(\omega+1)+6\omega+2}{\omega-1},~~~~~P_T=\frac{2\Lambda+8}{1-\omega}~,
\end{eqnarray}
and
\begin{eqnarray}\label{e5-11b}
C=\frac{\Lambda(\omega+1)-6\omega-2}{\omega-1},~~~~~P_T=\frac{8-2\Lambda}{\omega-1}~,
\end{eqnarray}respectively. As a result, putting the above expressions into the effective
potential, we arrive at
\begin{eqnarray}\label{e5-10}
U''(a_s)=\frac{ a_s^{6 \omega-6} (6 \omega-5) (6\omega-4)\big(\Lambda(1 + \omega)
\pm(6\omega+2)\big)\pm a_s^{3 \omega-3} (3 \omega-2) (3 \omega-1) (8\pm 2\Lambda)}
{12 (\omega-1)} \pm \nonumber\\
3 a_s^{6\omega-2}\omega(6 \omega-1)~,
\end{eqnarray} corresponding to the cases $k=\pm1$ respectively. The relevant stability
/instability regions within $(\omega-\Lambda)$ parameter space are depicted in the middle and
lower couple of plots in Fig. 7, respectively. So, we can argue that for both the closed and
open universes, whether $\Lambda$ is positive or negative, a full realization of the non-singular
EC scenario could be achievable. The important thing is that independent of the sign
of $\Lambda$, EC occurs in spatially closed universes by respecting the SEC, but in its
open counterpart it occurs only by violation of SEC.

\section{Conclusions}

\emph{``Emergent Cosmology''} paradigm is going to be a
successful non-singular alternative for cosmological inflation.
This scenario consists of existence and stability of static
solutions called Einstein static universe (ESU) along with the
issue of graceful transition to the expected thermal history of
the standard cosmology. In an attempt for realization of a non-singular
emergent cosmology, here we have presented a total action (\ref{e2-1})
from a perfect fluid scalar-metric model universe with a perfect fluid
plus a cosmological constant $\Lambda$ as a distinct sources. We have
proceeded by obtaining the classical equations of motion using a super
Hamiltonian approach within the Schut'z representation. The central idea
in study of EC paradigm is \emph{``beginning with no beginning''}.
Namely it contains the idea that our current universe may has been appeared
from a primordial rigid state past eternally known as Einstein static
universe (ESU) instead of initial Big-Bang singularity. In this paper, we
have studied in details this pre-inflation scenario within the context of
the above mentioned model universe by using the \emph{``linear stability theory''}
as well as \emph{``effective potential formalism''} by concerning on the analysis of the
static solutions. This dynamical system analysis within the mentioned methods consists of
two cases: $\Lambda=0$ and $\Lambda\neq0$.

By focusing on the first order dynamical system,
We have shown that in the absence of the cosmological
constant, $\Lambda=0$, in the case of fixing some
values for the free parameters $C$ and $p_{T}$ of the
underlying cosmological model and also by imposing
some constraints on the equation of state (EoS) parameter
$\omega$, a permanent ESU is realizable for both flat
and closed universes. As listed in Table 1 for the cases
of flat and closed universes, the EoS parameter $\omega$
satisfies the conditions $\omega=1$ and $\frac{1}{3}
\leq\omega<1$, respectively. However, we found that just
for the latter case, the conditions provide a graceful
exit from the frozen state towards the cosmological inflation
and then beginning of the standard thermal history. To be more
concrete, Fig. 1 (the left panel) obviously shows that the sign of the eigenvalue
$\lambda^2$ switches from negative to the positive one as
the EoS parameter $\omega$  reduces slowly with the cosmic time.
This means that the universe eternally shall not be caught in
the initial static state so that finally joins the standard
cosmology and in this way the Big-Bang singularity issue could be
resolved.

By following a first order dynamical system analysis,
regarding the contribution of $\Lambda$ term
in the matter sector of the action (\ref{e2-1}) in addition to the
perfect fluid, we have obtained further and more severe stability
conditions for ESU in comparison with $\Lambda=0$ case,
see Table 2. Here the stable ESU
can be realized in closed and open underlying model
universes by setting some admitted values for the
free parameters $C$ and $P_T$ as well as $\omega$.
Our analysis has revealed that in the
presence of $\Lambda$, to reach a sustainable
ESU we only need a tiny proportion of the cosmological constant,
that is, $|\Lambda|\ll1$. Nonetheless, we found that
all cases listed in Table 2 cannot result
in a graceful realization of the EC paradigm. Rather, only
for the third case of a closed spatial geometry an emergent
cosmology happens gracefully.
Although this result, in terms of the type of spatial
geometry, is similar to our former case (with $\Lambda=0$),
here the admitted range of EoS parameter $\omega$ is wider,
that is, $-\frac{1}{3}<\omega<1$.\\

For an EC paradigm to be a successful scenario, existence of a future de Sitter (dS) global attractor is essential. For this purpose,
we have extended our dynamical system analysis into a phase
space with infinite extension. Our analysis shows that in the presence
of $\Lambda$ under certain constraints on $(p_T-\Lambda)$ parameter space with non-flat spatial geometries, there are
dS global attractors.\\

Concerning on the effective potential formalism we have applied this formalism as
a another approach to investigate the dynamical properties of the EC scenario
within the cosmological toy model at hand. Indeed here ESU is realized by translation of
the required conditions within the effective potential formalism through derivation of the relevant
extremum points. By doing this analysis in the absence an also presence of $\Lambda$, we have
derived a new ESU which if certain constraints on the $\omega$ are met, then it can result
in a full realization of EC scenario. Of course, some of these constraints may cause an explicit
violation of SEC. However, the strength of this analysis within the framework of the underlying
cosmological model is that relevant outcomes are independent of fixing some values for free parameters
$C$ and $P_T$ of the model, unlike the first method. Figs 6 and 7 qualitatively reflect
the main results of this dynamical system analysis.\\

In summary and as our main conclusion, in this study we have shown that in
the absence or presence of the cosmological constant $\Lambda$, there is the
possibility of a graceful realization of Emergent Cosmology (and therefore a
resolution of the Big-Bang singularity problem) within a non-flat underlying
model universe with a perfect fluid supported by a variety of matter sources
which some of them are committed to SEC and some other violate it.

\section*{Acknowledgments}
We would like to thank Luca Parisi and Babak Vakili for insightful discussions.
We appreciate an anonymous referee for his/her constructive comments
that have improved the quality of the manuscript. This work has been supported financially by Research
Institute for Astronomy and Astrophysics of Maragha (RIAAM) under research project number 1/5440-65.

\end{document}